\documentstyle[galley,epsf]{mn}

\begin{document}

\title {A Search for $^{6}$Li in Stars with Planets}

\author[Reddy et al.] 
        {Bacham E. Reddy,$^1$ David L. Lambert,$^1$ Chris Laws,$^2$
        \newauthor 
        Guillermo Gonzalez,$^3$ Kevin Covey $^2$ \\
        $^1$Department of Astronomy, University of Texas, Austin, Texas 78712 \\
        $^2$Department of Astronomy, University of Washington, Seattle, Washington 98195\\ 
        $^3$Department of Physics and Astronomy, Iowa State University, Ames, Iowa, 50011 \\  }


\pagerange{\pageref{firstpage}--\pageref{lastpage}}

\maketitle

\label{firstpage}

\begin{abstract}
Using very high-resolution (R$\sim$ 125,000) and high quality (S/N $\geq$ 350) spectra,
we have searched for $^{6}$Li in stars hosting 
extra-solar planets.
From detailed profile-fitting
of the Li\,{\sc i} resonance line at 6707.7~\AA, we find no significant
amount of
$^{6}$Li relative to the $^{7}$Li for any of 8
planet bearing stars ($^{6}$Li/$^{7}$Li $\leq$ 0.0 - 0.03 ) with a strong Li\,{\sc i} lines.
In particular, we do not confirm the presence of $^{6}$Li with $^{6}$Li/$^{7}$Li = 0.13 
reported by Israelian et al.(2001) for HD\,82943, a star with two known planets. Several
of the 8 stars plus HD\,219542 A, the planet-less primary of a binary, have been identified
in the literature as possible recipients of accreted terrestrial material. For all of the planet-hosting
stars and an additional 5 planet-less stars, we find no $^{6}$Li. 

\end{abstract}


\section{Introduction}
Observers studying the abundance of lithium in stars continue to
uncover novel results. A recent novelty
concerning main sequence stars was reported by Israelian et al. (2001)  from
measurements of the lithium isotopic ratio for HD 82943, a star known to
host  giant planets, and HD 91889, a star similar to HD 82943 but lacking
orbiting planets. For HD 91889, analysis of the 6707 \AA\ Li\,{\sc i} resonance
line showed a complete absence of $^6$Li ($^6$Li/$^7$Li = $-$0.002 $\pm$
0.006, by number) in agreement with theoretical expectation that $^6$Li but
not $^7$Li is
thoroughly destroyed by protons in the pre-main sequence phase when the star
possesses a deep convective envelope. The abundance of lithium, that is
$^7$Li, at $\log\epsilon$(Li) = 2.52 is  a typical value for a late F-type main
sequence star. In striking contrast, HD 82943, a star with a similar total
lithium abundance as HD 91889, showed a surprising amount of $^6$Li: the
ratio determined was $^6$Li/$^7$Li = 0.126 $\pm$ 0.014. This extraordinary
amount of $^6$Li was cited by Israelian et al. as probable `evidence for a
planet (or planets) having been engulfed by the parent star'. Since lithium
$^6$Li and $^7$Li -- would presumably have been preserved in the unfortunate
planets, their ingestion  added $^6$Li (and $^7$Li) to the atmosphere
cleansed of $^6$Li in the pre-main sequence phase.

In this paper, we report measurements of the lithium isotopic abundance
ratio for 8 stars (including HD 82943) known to host extra-solar planets, and
for several similar stars not known to have extra-solar
planets.  Our principal result is that
$^6$Li is not detected in HD\,82943 or other stars. The next
section  describing our observations is followed by a thorough
description of our method of analysis.  We use HD 82943 as one of the stars with
which to illustrate our method.
After presenting our results for the $^6$Li/$^7$Li ratio, we
discuss the implications of the non-detections of
$^6$Li, as well as offering a few remarks on the lithium abundance in the
atmosphere of stars which host extra-solar planets.

\begin{table*}
\centering
\begin{minipage}{140mm}
\caption{Atmospheric parameters and observational details of the program stars.
           The absolute magnitudes, ages, and the masses were estimated using Hipparcos parallaxes and
            the apparent magnitudes combined with the isochrones and the evolutionary tracks 
            computed by Girardi et al. (2000).}
  \begin{tabular}{@{}llccrcccl@{}}
\hline \hline
   Star     &  $T_{\rm {eff}}$  & log $g$  & $\xi_{\rm {t}}$  & [Fe/H]  & M$_{\rm {v}}$  & Age  & M/M$_{\sun}$ &Ref\footnote{
References providing the adopted atmospheric parameters:- 
CYQ: Chen et al. (2000); GG: Gonzalez et al. (2001); IS: Israelian et al. (2001); KF: Fuhrmann (1998); 
LG: Laws \& Gonzalez (2001); PS: present study; RGG: Gratton et al. (2001);
SIM: Santos et al. (2000). } \\
            &  (K)              & (cm s$^{-2})$  & (km s$^{-1})$  &       &                 & (Gyrs)& &  \\
\hline
HD\,8574        & 6200 & 4.20 &1.32 &   0.12  & 4.00$\pm$0.50\footnote{Values of M$_{v}$ and Age
are uncertain as stars lack either measured parallax or accurate visual magnitudes.}  & 2.0$\pm$2.0$^{b}$& 1.15 & PS \\
HD\,10697        & 5605 & 3.96 & 0.95 &   0.16  & 3.73$\pm$0.05 & 7.1$\pm$1.0& 1.10 & GG \\
HD\,52265        & 6189 & 4.40 & 1.30 &   0.28  & 4.06$\pm$0.05 & 1.0$\pm$0.5 &1.20 &GG  \\
HD\,75289        & 6140 & 4.51 & 1.47 &   0.28  & 4.05$\pm$0.04 & 1.0$\pm$0.5 &1.20 &SIM  \\
HD\,82943        & 6010 & 4.62 & 1.08 &   0.32  & 4.35$\pm$0.05 & 1.0$\pm$0.5 &1.00  &SIM \\
HD\,89744        & 6338 & 4.17 & 1.55 &   0.30  & 2.79$\pm$0.06 & 2.0$\pm$0.5  &1.40 & GG  \\
HD\,141937       & 6150 & 4.30 & 1.40 &   0.31  & 4.20$\pm$0.50$^{b}$ & 0.8$\pm$2.0$^{b}$  &1.22 & PS \\
16 Cyg B         &5685&4.26 & 0.80 &   0.07  & 4.55$\pm$0.02 & 9.0$\pm$1.0 &1.00 & LG \\
HD\,209458       & 6063 & 4.38 & 1.02 &   0.04  & 4.29$\pm$0.10 & 4.0$\pm$1.0 &1.10 &GG \\
HD\,219542~A\footnote{Stars not known to have planets.}    & 5989 & 4.37 & 1.20 &   0.29  & 4.50$\pm$0.50$^{b}$&  1.0$\pm$2.0$^{b}$            &1.15 & RGG \\
HD\,75332$^{c}$  & 6305 & 4.49 & 1.05 &   0.24  & 3.93$\pm$0.05 & 1.0$\pm$1.0 &1.25 & GG  \\
HD\,91889$^{c}$  & 6070 & 4.41 & 1.52 & $-$0.23  & 3.75$\pm$0.03 & 7.9$\pm$0.5 & 0.95 &IS \\
HD\,142373$^{c}$ & 5920 & 4.27 & 1.1  & $-$0.39 &  3.62 $\pm$0.02 & 5.0$\pm$0.5 &0.90 &CYQ \\
HD\,154417$^{c}$ & 5925 & 4.30 & 1.1  & $-$0.04 &  4.46$\pm$0.04 & 4.0$\pm$0.5 &1.05 & CYQ\\
HD\,187691$^{c}$ & 6088 & 4.07 & 1.35 &   0.07  & 3.66$\pm$0.03 & 4.0$\pm$0.5 &1.25 &KF \\
\hline
\end{tabular}
\end{minipage}
\end{table*}

\begin{figure*}
\epsfxsize=18truecm
\epsffile{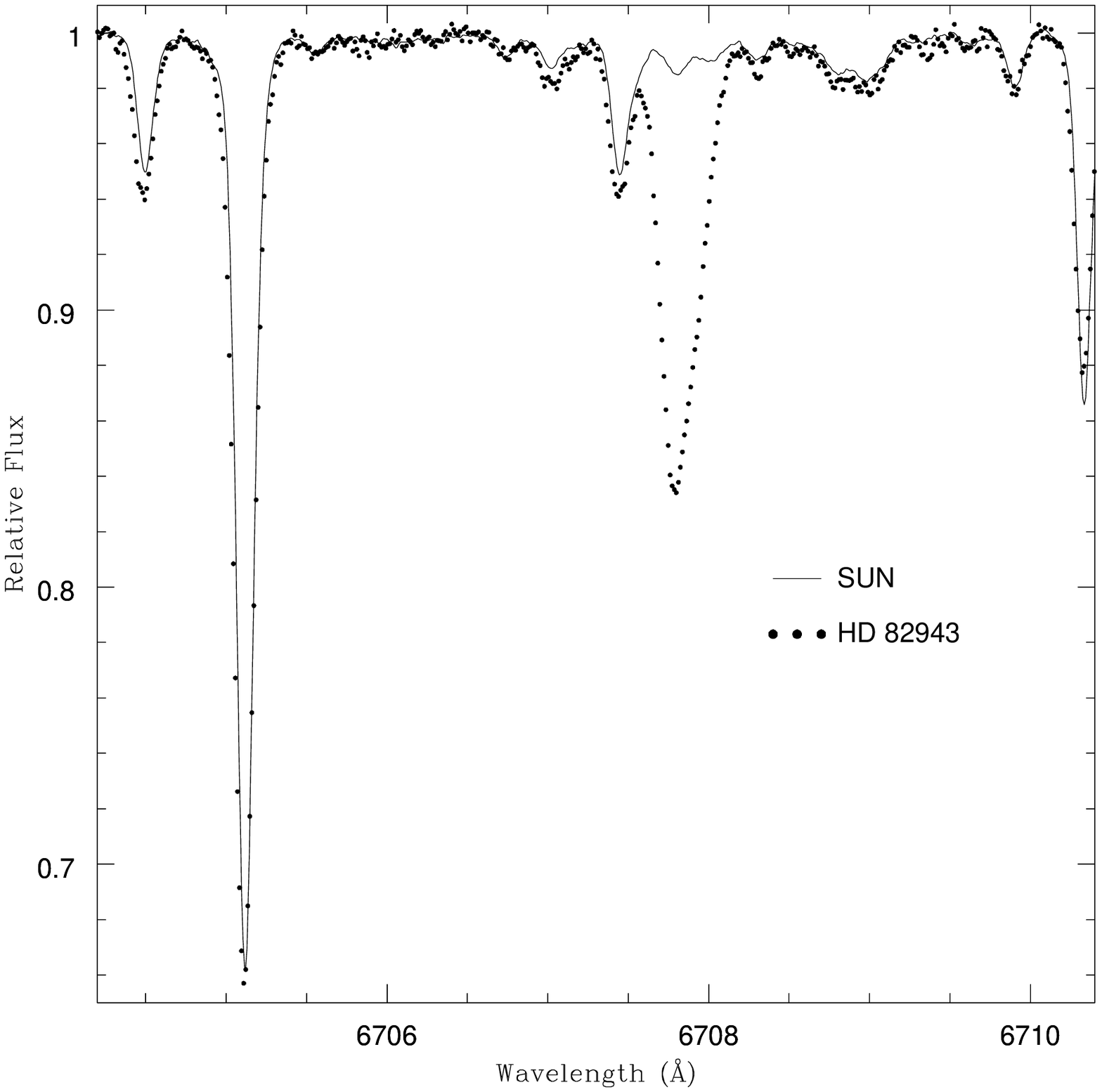}
\caption{ A comparison of the spectrum of HD 82943 and the Sun
from 6705 \AA\ to 6710.25 \AA. Note the similar strengths and widths of the
several Fe\,{\sc i} lines but the great difference in the
Li\,{\sc i} 6707.7 \AA\ feature.}
\end{figure*}
\section{Observations}

The observed stars listed in Table 1 fall into two
categories: stars with extra-solar planets, and similar stars for which
there are no reported planets. Chosen stars  have a Li\,{\sc i}
6707 \AA\ line of a strength enabling a meaningful limit (or detection) to be
set on the lithium isotopic ratio. Sharp-lined stars were  given priority.
The star 16 Cyg B, a host for planets but almost lacking in lithium, was observed in order to
check on lines that might blend with the Li\,{\sc i} 6707 \AA\ feature.
Comparison stars of similar atmospheric parameters were observed including
HD\,91889 with others chosen from surveys
for lithium abundances by Balachandran (1990) and Chen et al. (2000).

All of our observations were obtained with the coud\'{e} echelle
cross-dispersed spectrograph {\it 2dcoud\'{e}} at the W.J. McDonald
Observatory's 2.7m Harlan J. Smith reflector (Tull et al. 1995). The
spectrograph was used in its higher resolution mode.  The recorded spectrum
provides incomplete coverage of 12 orders. The order containing 6707 \AA\
covers 25~\AA. The resolving power was $R = \lambda/\Delta\lambda$ = 125,000, as
determined from the widths (FWHM) of thorium lines in the comparison
spectrum. The spectra are over-sampled: there are 4.5 pixels per resolution
element.

Several  exposures of a star were obtained with an exposure time of 15
to 20 minutes with total integration time of around  1 to 2 hours. 
A ThAr hollow cathode lamp was observed before and after a
sequence of
exposures on a star. Each image recorded on a Tektronix 2048 $\times$ 2048
CCD was processed  in the usual way using  IRAF subroutines to obtain a one
dimensional spectrum. The  wavelength calibration was applied to each individual
exposure and was accomplished using a third-order polynomial fit to a large
number of Th emission lines; this resulted in a typical rms accuracy of 2 m\AA.

Over the course of a sequence of exposures on a given star, the spectrum
drifted on the CCD owing to the change in the  Earth's motion, and to
instrumental drifts. By cross-correlation,  spectra were  shifted so that
they matched the first spectrum of a sequence. Continuum normalization of the
resultant spectrum was achieved by fitting a  sixth-order polynomial to each
order. The S/N per pixel in the continuum near 6707 \AA\ is listed in Table 2. As a
check on the continuum fitting, we compare the spectrum of HD 82943 near
6707 \AA\ with that of the solar spectrum published by Hinkle et al. (2000)
(Figure 1). It is seen that the line-free intervals in HD 82943 correspond
very closely to those in the solar spectrum; the rms differences of about
0.2$\%$ are consistent with the S/N ratio of our spectrum of HD 82943.

\section{Analysis: Preliminaries}

In stellar spectra the Li\,{\sc i} resonance doublet
with a
separation of  0.15 \AA\ between the stronger blue and weaker red
components
is not resolved into two  lines;
the stellar line is an asymmetric blend with a greater
width than other stellar lines of  comparable strength. Addition of $^6$Li
with its stronger line almost at the wavelength of the weaker $^7$Li line
and its weaker line placed at an additional 0.15 \AA\ to the red introduces a
wavelength shift of the blend and increases its asymmetry. Therefore, in order to
extract the isotopic ratio from a stellar line, it is necessary to establish
an accurate wavelength scale, and to quantify the various factors
responsible for the line broadening.

A necessary  condition for defining the zero point of
the wavelength scale is the
determination of the stellar radial velocity. But examination of our spectra
shows that it is not a sufficient condition because the radial velocity varies
with the strength of the measured line. This
is due to velocity shifts induced by the convective motions
(granulation) in the stellar atmosphere.
The conventional list of factors influencing the width of a weak line like
the 6707 \AA\ feature includes the projected rotational velocity ($v\sin i$),
the macroturbulence, the microturbulence, and the instrumental profile. It
is also necessary to consider the blends that contribute to the stellar 6707
\AA\ line.

Reduced spectra were analyzed independently at the University of Texas (UT) and the
University of Washington (UW). Results reported here for the full sample are those
from the UT analyses, but are confirmed by the UW results. Some differences
in the two approaches are noted later.

\begin{figure*}
\epsfxsize=18truecm
\epsffile{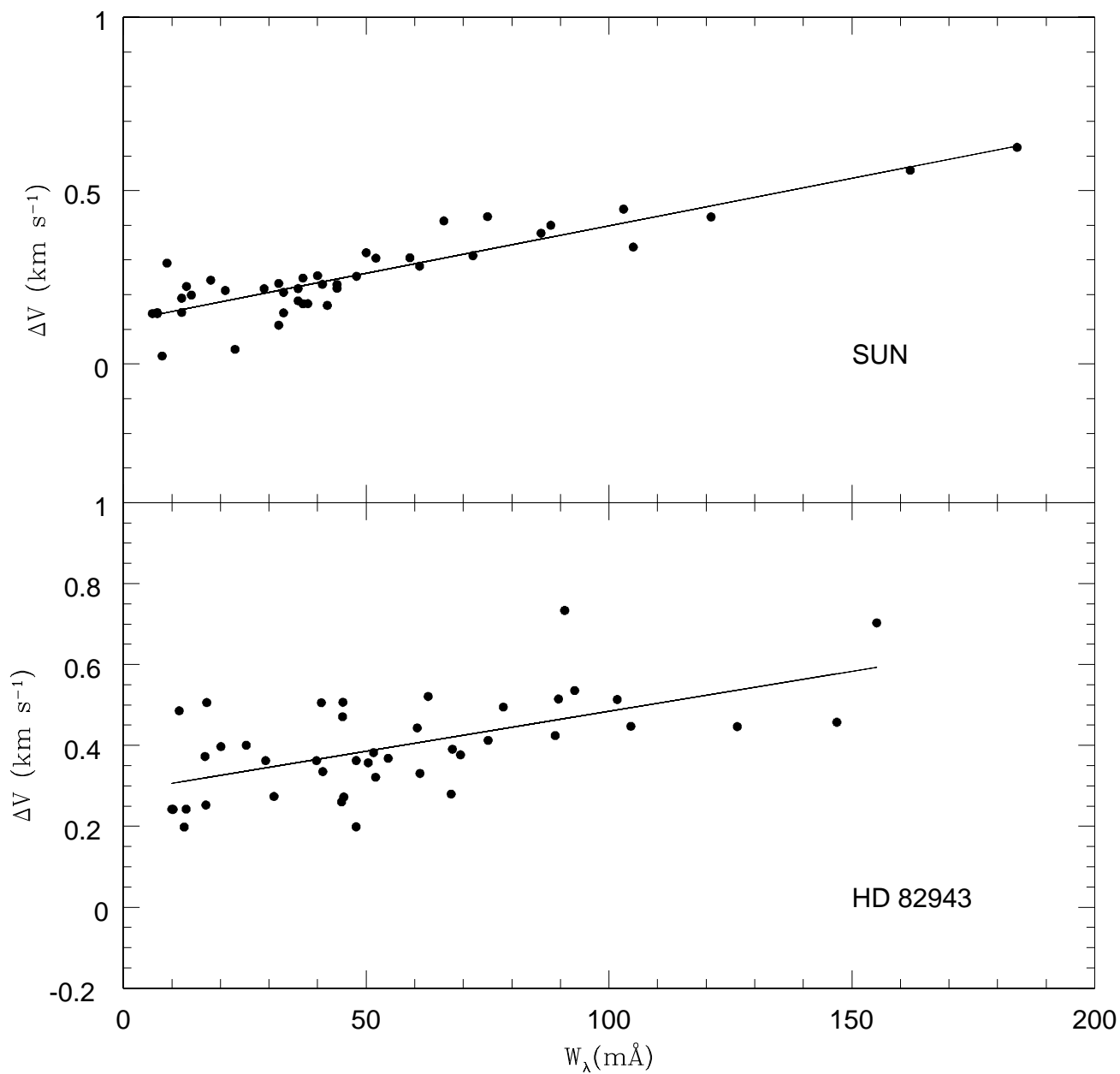}
\caption{Radial velocity differences for Fe\,{\sc i} lines as
a function of equivalent width for HD 82943 (bottom panel) and the Sun
(top panel). The trend to increasing redshift with increasing
equivalent width is similar in the two cases. The solid line
is the least squares fit to the data points for HD\,82943. The same
is shown in the upper panel for the Sun.}
\end{figure*}
\subsection{Radial Velocity and Convective Velocity Shifts}

The stellar radial velocity was measured for a sample of unblended Fe\,{\sc
i} lines using the  accurate laboratory wavelengths  reported by Nave et al.
(1994). There is a clear trend of velocity with equivalent width
($W_\lambda$) (Figure 2) in that the weaker lines are blue-shifted with
respect to the stronger lines. This is a well known phenomenon attributed to
granulation -- see the set of extensive measurements
 on Fe\,{\sc i} lines in the Sun and
Procyon reported by Allende Prieto \& Garc\'{\i}a L\'{o}pez (1998) and
Allende Prieto et al. (2002), respectively. The parallel with the Sun is
almost exact, as might be expected from Figure 1. Velocities of the same
Fe\,{\sc i} lines have been measured from Hinkle et al.'s (2000) solar
atlas (Figure 2).

Convective  shifts are  expected to depend on the atom or ion providing
the line and on the line's excitation potential as well as its strength. In
order to estimate the shift that might be expected of the Li\,{\sc i}
resonance line, we measured the solar shifts for lines in the red due
to  low excitation lines of Na\,{\sc i}, Al\,{\sc i}, and Ca\,{\sc i}:
there are few of these lines in the limited bandpass of our stellar spectra.
 To a satisfactory precision, the velocity shifts of
the additional lines follow the dependence of the Fe\,{\sc i} line shifts
on $W_\lambda$.
Since the Fe\,{\sc i} shifts for HD 82943 match well the solar
shifts, we assume that the same equality holds for the Na\,{\sc i}, Al\,{\sc
i}, and Ca\,{\sc i} lines in this star. It should be noted that
expected differential velocity shifts between the Li\,{\sc i} feature and metal lines
in the same order are small ($\sim 0.2$ km s$^{-1}$ or less)
relative to the isotopic shift of about 6.7 km
s$^{-1}$ but it is, as we note below, an observable effect.

Granulation also renders stellar lines asymmetric. The line bisector is the
usual way to express the asymmetry. Such an asymmetry is  measurable
for  the stellar lines on our spectra.
The bisector for a line in our program stars is  similar to
that of the same line in the solar spectrum. The asymmetry is very
much smaller than the isotopic shift of the lithium line.

\subsection{Stellar Atmospheric Parameters}

Calculation of synthetic spectra requires a model atmosphere, and a line
list. A model atmosphere was chosen from the grid
provided by Kurucz (1995) according to the parameters $T_{\rm eff}$, log $g$,
and [Fe/H] derived from published abundance analyses (see references
given in Table 1. In this analysis we used an updated 
spectral analysis code MOOG (Sneden, 1973).
The lithium isotopic analysis is insensitive to the
particular choice of parameters. This and the fact that the published
analyses are comprehensive mean that it is not necessary to derive afresh
these parameters.
For two stars with extra-solar planets, HD\,8574 and HD\,141937, 
with atmospheric parameters unavailable in the literature, the parameters
$T_{\rm eff}$, log $g$, $\xi_{t}$, and [Fe/H] were derived 
using 50 well defined Fe\,{\sc i} and 4 Fe\,{\sc ii} lines with the required oscillator
strengths ($gf$-values) for Fe\,{\sc i} lines taken from Nave et al. (1994) and 
from Giridhar \& Ferro (1995) for Fe\,{\sc ii} lines.
We estimate uncertainties of about  $\pm$150~K in $T_{\rm {eff}}$, $\pm$0.25 in log $g$,
0.25 km s$^{-1}$ in $\xi_{t}$, and $\pm$0.10 in [Fe/H].

To match  synthetic  to observed spectra requires information on the
various atmospheric
effects that broaden the stellar profiles. The UT analysis
considered three Fe\,{\sc i} lines in the same echelle order as the Li\,{\sc
i} feature and of a similar strength to it. They are the lines at 6703.5,
6705.1, and 6713.7 \AA. Synthetic profiles were computed using the appropriate
model atmosphere,  the iron abundance, and the
 microturbulence
 to match a Fe\,{\sc i} line's equivalent width. Two independent schemes were then
employed to fit the synthetic line to the observed line
profile.
In the first scheme, the synthetic profile was
broadened by a Gaussian modeled macroturbulence defined by
$\Gamma_{g}$, representing the combined effects of stellar rotation, macroturbulence,
and the instrumental
profile as provided by the thorium lines in the comparison spectrum.
In the second scheme, macroturbulence and rotational broadening were treated
as different sources of broadening.

In the first scheme, the $\Gamma_{g}$ was varied to match the observed
profile of a selected Fe\,{\sc i} line
with the best fit judged by the reduced $\chi^2_{r}$ where

\begin{equation}
\centering
\chi^2_{r} = \frac{1}{d}\sum^{n}_{1}\frac{(O_i - P_i)^2}{\sigma^2}
\end{equation}

where $O_i$ and $P_i$ respectively are the observed  and predicted relative
fluxes at data point $i$  across the line profile, $\sigma$ is the rms error
of the continuum, $d = n-c$ (= 22 for Fe\,{\sc i} and 35 for Li\,{\sc i},
typically) is the number of degrees of freedom in the fit, $n$ is the
number of observed data points involved in the
fit, and $c$ is the number of parameters to be determined from
the computed line profile
fit. Here, $c$ is   3 for Fe\,{\sc i} lines  (i.e., abundance, radial velocity,
and the parameter $\Gamma_{g}$),
and 4 for the Li\,{\sc i} line with the isotopic $^{6}$Li/$^{7}$Li ratio
as the additional parameter. The run of $\chi^{2}$ with $\Gamma_{g}$ is
plotted and the minimum value found.

\begin{figure*}
\centering
\epsfxsize=18truecm
\epsffile{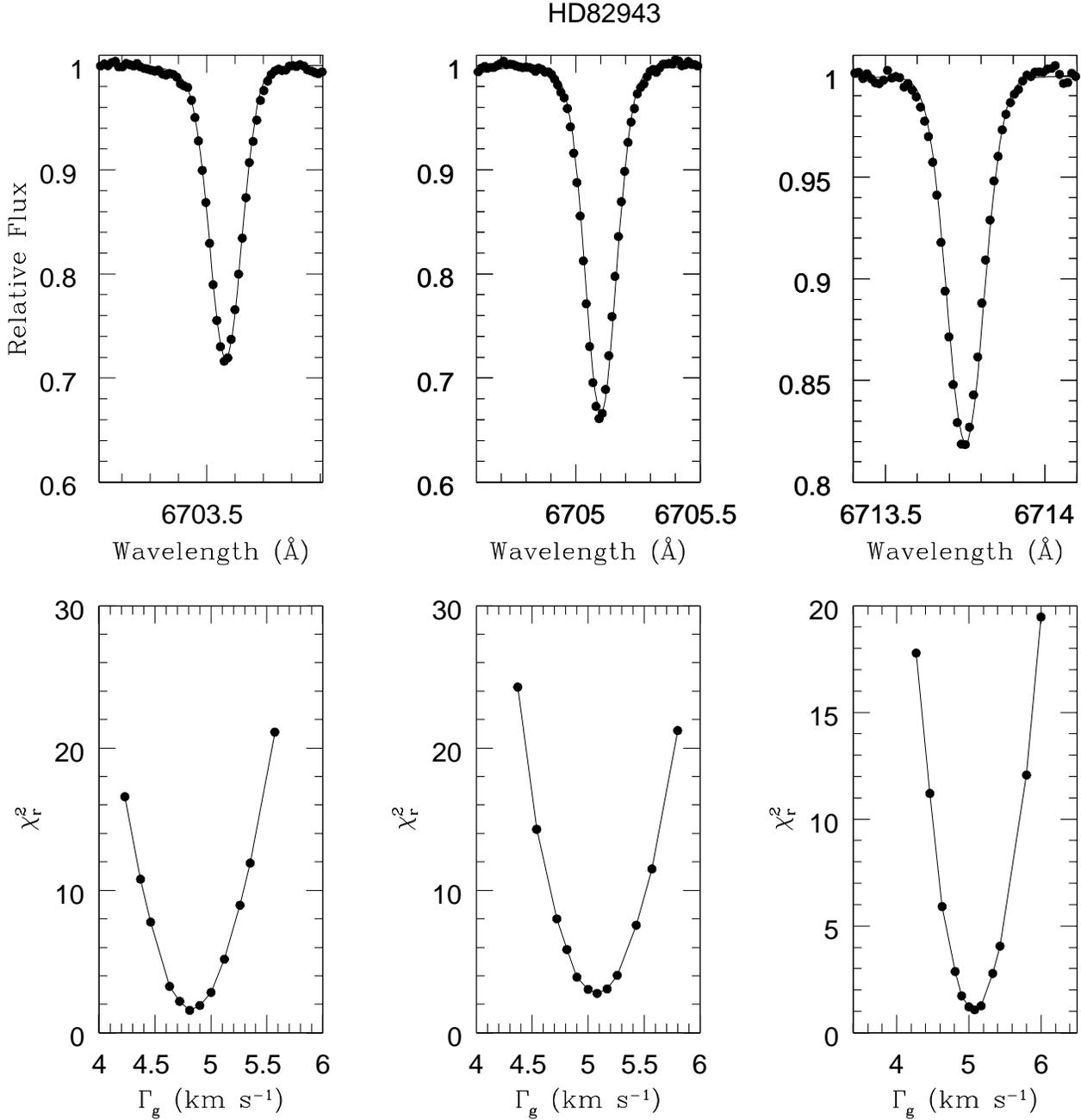}
\caption{Three Fe\,{\sc i} lines in the spectrum of HD 82943
showing the fit of the  synthetic profile for the optimum
value of the Gaussian 
macroturbulence $\Gamma_{g}$. The lower panels show the run of
$\chi^2_{\rm r}$ with $\Gamma_{g}$.}
\end{figure*}

\begin{figure*}
\epsfxsize=18truecm
\epsffile{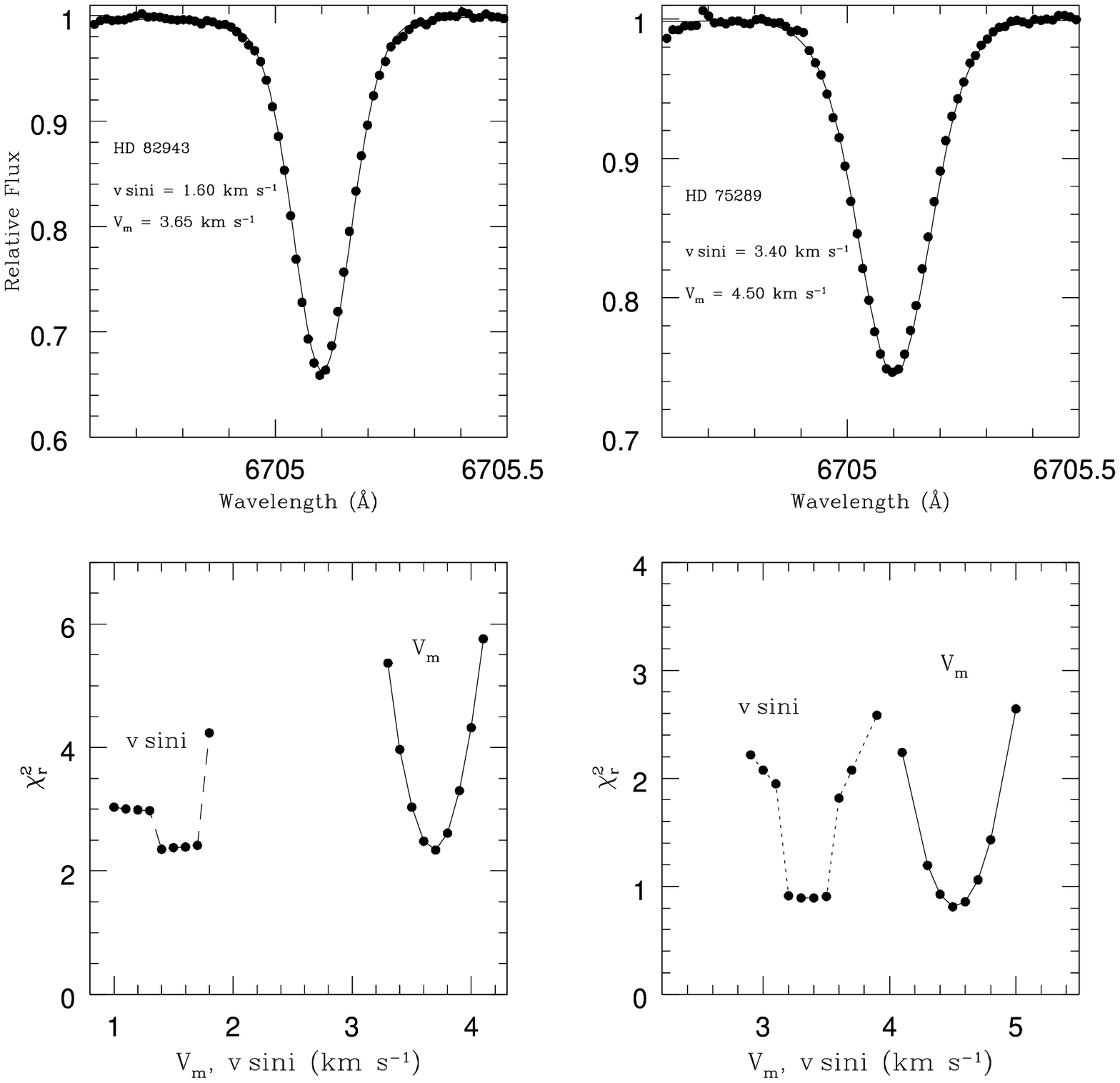}
\caption{Determination of $v\sin i$ and $V_{\rm m}$
for HD\,82943 and HD\,75289}
\end{figure*}

The agreement between the predicted and the observed profiles
is quantified with the $\chi^{2}$-test. The fit is considered
very good when the $\chi^{2}_{r}$ $\simeq$ 1 for the best fit
(Bevington \& Robinson 1992).
For HD 82943, Figure 3 shows the three Fe\,{\sc i} profiles and the run of
$\chi^2_{r}$ as a function of $\Gamma_{g}$ in the first scheme.
Two of the lines are fit with a minimum $\chi^2_r$ $\simeq 1$, and the
third with a slightly higher value.
The three results  are consistent with
a mean value $\Gamma_{g}$ = 5.03$\pm$0.12 km s$^{-1}$. Such a determination ignores
the small line asymmetry.
Rotational and instrumental broadening are subsumed into $\Gamma_{g}$.

In the second scheme, all stars were also analyzed with $v\sin i$ and $V_m$ as free parameters.
The $v\sin i$ and $V_m$ values are determined for a given Fe abundance (determined from
equivalent widths) such that the computed profiles best fit (i.e minimum $\chi^{2}_{\rm r}$ values) the observed Fe profiles.
Figure 4 shows the results for an Fe\,{\sc i} line for HD 82943 along with a similar star
HD 75289.
The three Fe\,{\sc i} lines were fitted in this way.
The quality of the fits to the observed profile is quite similar
to that obtained just using $\Gamma_{g}$.
Results for $\Gamma_{g}$, $V_m$ and $v\sin i$ are given in Table 2 for all 
the program stars. 

In the UW analysis, the Fe\,{\sc i} lines at 5852.2, 6330.9, 6703.5, and 6705.1, and 6715.4 \AA\
were employed to determine the line broadening parameters $v\sin i$ and $V_{m}$.
Instrumental broadening was handled in the same manner as the UT group and the
limb darkening coefficients were estimated using the work of Gray (1992).
A linelist for 1~\AA\ regions surrounding each line was prepared via comparison of synthetic
profiles generated using the MOOG spectral synthesis code (Sneden, 1973) at 0.001 \AA\ resolution
with both the Solar Flux Atlas (Kurucz et al. 1984) and spectra of 16 Cyg B. As in similar
prior work by Gonzalez (1998) and Takeda (1995), we discovered that in order to achieve good fits
to these Fe\,{\sc i} features using accurate atmospheric parameters and accepted values
for solar rotational and macroturbulent velocities, it was necessary to both
reduce the microturbulent velocity, $\xi_{\rm t}$, by a factor of 0.4 and allow for a 0.03 - 0.05 dex
offset in [Fe/H]. Applying this correction factor to $\xi_{\rm t}$ results in a more
accurate $v\sin i$. However, adopting a standard value of $\xi_{\rm t}$ would still
result in a good fit to the line profiles.
Synthetic profiles were calculated for each line using Kurucz (1995) model atmospheres, selected
according to published values of $T_{\rm eff}$, log $g$, and [Fe/H] (Table 1); accompanying
estimates of $\xi_{\rm t}$ were adjusted in a manner consistent with that performed to fit
the solar spectra. In each case, the precise placement of the continuum level was set to
minimize the reduced $\chi^{2}_{\rm r}$ values of fits to adjacent continuum regions, and the velocity offset $\Delta v$
was adjusted to minimize the difference between the centroids of the observed and synthesized Fe lines.

The UW analysis is based on 10-20 thousand synthesized profiles for each line at increments of 0.05 km s$^{-1}$ in
both $v\sin i$ and $V_{m}$ and 0.01 dex in [Fe/H]. Values of $\chi^{2}_{\rm r}$ were calculated for each
fit, resulting in 2-dimensional grids of $\chi^{2}_{\rm r}$ as a function of $v\sin i$ and $V_{m}$ for
4-7 values of [Fe/H]. 
For each
Fe\,{\sc i} line, we adopt those values of $v\sin i$ and $V_{m}$ which yielded the minimum values
of $\chi^{2}_{\rm r}$ amongst all competing fits. We take as our final solution for each star
the average of the best-fit broadening parameters found from each line, and quote the standard deviation
of their mean as an estimate of uncertainty.
The final resulting broadening parameters for the UW group
are given in Table~2.

\begin{table*}
\centering
\begin{minipage}{140mm}
\caption{ Lithium isotopic $^{6}$Li/$^{7}$Li ratio results
derived from 6707~\AA\ Li-profile. One-$\sigma$ errors are quoted.}
\begin{tabular}{lclllccllll}
\hline
\hline
Star      & S/N & $\Gamma_{g}$ &  v\,sini   &  $V_{m}$  & v\,sini\footnote{Values
independently derived by UW group.}
& $V_{m}^{a}$ & $\log\epsilon$(Li)\footnote{Values
computed using the single Gaussian broadening parameter $\Gamma_{g}$.} & $^{6}$Li/$^{7}$Li$^{b}$ 
      & $\chi^{2}_{min}$$^{b}$\\
      &     & (km s$^{-1}$)   &  (km s$^{-1}$)  &  (km s$^{-1}$) & (km s$^{-1}$) & (km s$^{-1}$) &(dex) &    &   & \\
\hline
HD\,8574  & 300 &7.22$\pm$0.16& 3.30$\pm$0.20& 4.65$\pm$0.10&  3.60$\pm$0.26 & 5.30$\pm$0.30  & 2.71$\pm$0.05 & 0.03$\pm$0.05 & 1.02 \\ 
HD\,10697 & 300 &5.27$\pm$0.10&1.20$\pm$0.30 & 4.30$\pm$0.15&  1.61$\pm$0.19 & 4.04$\pm$0.18  & 1.91$\pm$0.03 & 0.00$\pm$0.06 & 1.15 \\
HD\,52265 & 450 &7.39$\pm$0.08& 3.10$\pm$0.25 & 5.50$\pm$0.13& 3.88$\pm$0.11&4.81$\pm$0.16&2.76$\pm$0.01 & 0.01$\pm$0.03 &1.05\\
HD\,75289 & 350 &7.08$\pm$0.10& 3.40$\pm$0.25 & 4.50$\pm$0.10& 4.17$\pm$0.25&4.15$\pm$0.30&2.77$\pm$0.02 & 0.02$\pm$0.03 & 0.90 \\
HD\,82943 & 450 &5.03$\pm$0.12& 1.60$\pm$0.20 & 3.65$\pm$0.10& 2.09$\pm$0.19&3.73$\pm$0.13&2.43$\pm$0.02 & 0.00$\pm$0.02 & 1.06 \\
HD\,82943\footnote{Two spectra of HD\,82943 were obtained and independently analysed.} & 
530 &  5.00$\pm$0.14 &1.55$\pm$0.13  & 3.73$\pm$0.15  &  ...      & ...  & 2.43$\pm$0.02 & 0.01$\pm$0.03 & 1.20 \\
HD\,89744 & 550 &13.54$\pm$0.20& 7.60$\pm$0.28& 8.00$\pm$0.20&8.02$\pm$0.13&7.71$\pm$0.32 &2.11$\pm$0.01 & 0.00$\pm$0.03 & 1.20 \\
HD\,141937& 380 & 5.14$\pm$0.25& 1.15$\pm$0.30& 3.85$\pm$0.15& 1.89$\pm$0.26 & 4.00$\pm$0.29   & 2.55$\pm$0.04 & 0.01$\pm$0.03 & 1.12 \\
16 Cyg B  & 600 &4.79$\pm$0.10& 1.60$\pm$0.25&  3.30$\pm$0.18&1.64$\pm$0.16& 3.49$\pm$0.16 &0.71$\pm$0.02&0.00$\pm$0.03 & 1.30 \\
HD\,209458& 450 &7.24$\pm$0.05& 3.33$\pm$0.15& 4.83$\pm$0.18 &3.57$\pm$0.11& 5.24$\pm$0.22 & 2.70$\pm$0.02&0.00$\pm$0.03& 0.96 \\
HD\,219542~A& 300 &4.50$\pm$0.20 & 1.55$\pm$0.25& 3.55$\pm$0.10& 2.24$\pm$0.16& 3.66$\pm$0.12 &2.26$\pm$0.03 & 0.03$\pm$0.04 & 1.06\\
HD\,75332 & 650 &12.62$\pm$0.21& 8.05$\pm$0.20& 6.78$\pm$0.23 &8.20$\pm$0.06&6.60$\pm$0.16 &3.18$\pm$0.01 & 0.00$\pm$0.02 & 1.05 \\
HD\,91889 & 600 &6.63$\pm$0.12&  2.20$\pm$0.30& 5.15$\pm$0.16 &  2.89$\pm$0.26 & 4.49$\pm$0.14 & 2.46$\pm$0.02 & 0.00$\pm$0.02 & 0.98 \\
HD\,142373& 700 &5.12$\pm$0.13 & 1.05$\pm$0.20 & 4.28$\pm$0.15 &  2.21$\pm$0.51 & 3.67$\pm$0.46 & 2.52$\pm$0.01 & 0.00$\pm$0.01 & 1.36\\
HD\,154417& 600 &8.68$\pm$0.18 & 3.95$\pm$0.32 &  6.35$\pm$0.18 & 4.90$\pm$0.06 & 5.10$\pm$0.12 &2.68$\pm$0.01 & 0.00$\pm$0.01& 1.18\\
HD\,187691& 700 &6.40$\pm$0.08 & 1.90$\pm$0.20& 5.25$\pm$0.25&  2.88$\pm$0.12 & 5.20$\pm$0.18 &2.56$\pm$0.01 & 0.03$\pm$0.02 & 0.95\\
\hline
\end{tabular}
\end{minipage}
\end{table*}

\subsection{The Line List around 6707 \AA.}

Atomic data - wavelengths and $gf$-values - for the Li\,{\sc i}
6707 \AA\ doublet and its isotopic and hyperfine splittings are of
very high quality (Smith, Lambert, \& Nissen 1998; Hobbs, Thorburn, \&
Rebull 1999). For the UT line list, 
potential blending atomic and molecular (CN) lines were
identified from various sources: Kurucz (1995), Lambert, Smith, \& Heath (1993),
Brault \& M\"{u}ller (1975), and Nave et al. (1994). The UT line list
was tested and refined by matching the solar spectrum (Hinkle et al.
2000), and the spectrum of 16 Cyg B. Figure 5 shows the
fit to the solar and 16 Cyg B spectra suggesting the line list lacks a line
at about 6708.0~\AA. In their detailed analysis of the solar
spectrum around 6707 \AA, Brault \& M\"{u}ller noted the presence of
weak unidentified lines at 6708.025 \AA\ and 6708.110 \AA. Unfortunately,
these lines straddle the weaker (red) line of the $^6$Li doublet, and
should, therefore, be included in the line list. On the basis of
laboratory spectra available to them, Brault \& M\"{u}ller
concluded that these unidentified lines were not due to either
CN or Fe\,{\sc i}.

\begin{figure*}
\epsfxsize=18truecm
\epsffile{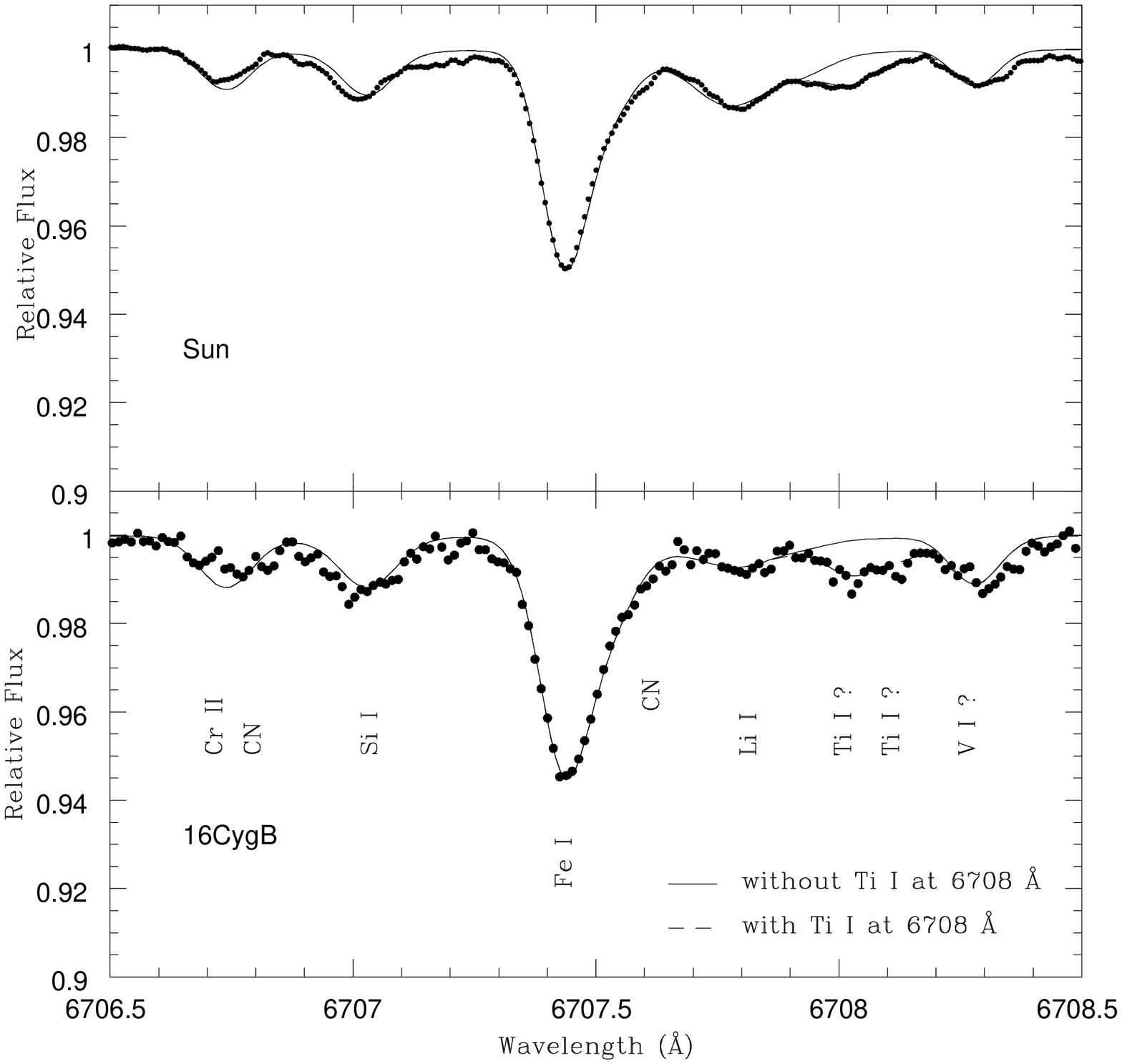}
\caption{Observed and synthetic spectra around the Li\,{\sc i} 6707 \AA\
feature for the Sun and 16 Cyg B. The contribution of the unidentified lines
near the weaker $^6$Li line is given by the difference between the dashed and
solid lines between 6707.9 \AA\ and 6708.2 \AA.}
\end{figure*}

In earlier studies of the 6707 \AA\ line (e.g., Nissen et al. 1999),
the presence of
the unidentified lines was noted, but ignored as their contribution is negligible
for metal-poor dwarfs. However,
in F-G main sequence stars of approximately solar metallicity,
this pair of lines will significantly affect the red wing of the
6707 \AA\ profile. 
King et al. (1997) noted the effect of these unidentified lines on the solar $^{6}$Li/$^{7}$Li ratio.
Treating these missing lines as $^{6}$Li components they obtain $^{6}$Li/$^{7}$Li = 0.10
for the Sun. This high Li isotopic ratio for the sun is very unlikely for
obvious reasons, and they attributed the lines to
Si~I.
In the solar spectrum, the 6708.025 \AA\ and
6708.110 \AA\ lines
have  equivalent widths of $W_\lambda = 1.4$ m\AA, and $<$ 0.5 m\AA,
respectively.
Our analyses of the solar and 16 Cyg B spectra give lithium ($^{7}$Li) abundances consistent
with previous analyses. For the Sun, our line list with the Kurucz
solar model gives $\log\epsilon$(Li) = 0.98 and for 16 Cyg B we obtain $\log\epsilon$(Li) = 0.71 which
are close to the values of $\log\epsilon$(Li) = 1.05$\pm$0.06 and $\log\epsilon$(Li) $\leq$0.60, respectively, derived by
King et al. (1997). 

Our analysis of selected stars was made first without the unidentified
lines in the line list. Then, we repeated the analysis on the
assumption that they were
neutral Ti lines with a lower excitation potential of 1.88 eV
and derived their $gf$-values  from a fit to  the solar
spectrum. As a check, we synthesised the spectrum of 16 Cyg B and
found that the (unidentified) Ti\,{\sc i} lines matched the
observed feature well (Figure 5). The star 16 Cyg B is cooler than most of our
program stars and is very deficient in lithium (King et al. 1997).

Figure~5 shows that while the $^{7}$Li feature
gets weaker, 
the missing feature at 6708.0~\AA\ is a little stronger in 16 Cyg B
than in the solar spectrum. 
We have varied the assumed excitation potential of the Ti\,{\sc i} lines, and tested
to changes in $T_{\rm eff}$.
We found a Ti\,{\sc i} line with a LEP of 1.88 eV
varies with $T_{\rm eff}$ by around 0.08~dex/100~K, and for a 
LEP of 0.0 eV, the
Ti\,{\sc i} line strength varies by 0.12~dex/100~K. Thus, by adopting 1.88~eV lines
of neutral Ti for the unidentifed lines, instead of 0.0 eV lines, we overestimate their
strength by around 0.10~dex for stars of $T_{\rm eff}$ = 6400~K, the hottest
stars in our sample. Higher LEP lines of Ti and lines of higher ionisation potential elements like Si and Mg
also investigated. Within our $T_{\rm eff}$ range, the $T_{\rm eff}$ effect on strength of these lines
does not have noticeble effect on $^{6}$Li/$^{7}$Li ratios.
Given that our programme stars are quite similar to the Sun,
identification of the pair as
Ti\,{\sc i} lines  is not critical to the analysis, but their inclusion is critical. 
The final line list is given in Table~3.

\begin{table}
\centering
\caption{Adopted linelist at the vicinity of the Li 6707\AA\ profile}
\begin{tabular}{llll}
\hline
\hline
$\lambda$  & Element     & LEP  & log $gf$      \\
 (\AA)     &             & ($eV$) &  ($dex$)     \\
\hline
  6707.381 & CN          & 1.83& -2.170   \\
  6707.433 & Fe\,{\sc i} & 4.61& -2.283   \\
  6707.450 & Sm\,{\sc ii}& 0.93& -1.040   \\
  6707.464 & CN          & 0.79& -3.012   \\
  6707.521 & CN          & 2.17& -1.428   \\
  6707.529 & CN          & 0.96& -1.609   \\
  6707.529 & CN          & 2.01& -1.785   \\
  6707.529 & CN          & 2.02& -1.785   \\
  6707.563 & V\,{\sc i}  & 2.74& -1.530 \\
  6707.644 & Cr\,{\sc i} & 4.21& -2.140 \\
  6707.740 & Ce\,{\sc ii}& 0.50& -3.810 \\
  6707.752 & Ti\,{\sc i} & 4.05& -2.654 \\
 6707.7561 &  $^{7}$Li   & 0.00& -0.428 \\
 6707.7682 &  $^{7}$Li   & 0.00& -0.206 \\
  6707.771 &  Ca\,{\sc i}& 5.80& -4.015 \\
  6707.816 & CN          & 1.21& -2.317 \\
 6707.9066 &  $^{7}$Li   & 0.00& -1.509 \\
 6707.9080 &  $^{7}$Li   & 0.00& -0.807 \\
 6707.9187 &  $^{7}$Li   & 0.00& -0.807 \\
 6707.9196 &  $^{6}$Li   & 0.00& -0.479 \\
 6707.9200 &  $^{7}$Li   & 0.00& -0.807 \\
 6707.9230 &  $^{6}$Li   & 0.00& -0.178 \\
  6708.025 &  Ti\,{\sc i}& 1.88& -2.252 \\
 6708.0728 &  $^{6}$Li   & 0.00& -0.303 \\
  6708.094 &  V\,{\sc i} & 1.22& -3.113 \\
  6708.125 &  Ti\,{\sc i}& 1.88& -2.886 \\
  6708.280 &  V\,{\sc i} & 1.22& -2.178 \\
  6708.375 & CN          & 2.10& -2.252 \\
\hline
\end{tabular}
\end{table}

\section{The $^6$L{\i}/$^7$L{\i} Ratio}

Addition of $^6$Li to the mix of lithium isotopes shifts the
6707 \AA\ line to the red and increases the line's asymmetry.
For the meteoritic isotopic ratio ($^{6}$Li/$^{7}$Li $\sim$ 0.10),
the  center-of-gravity of the profile shifts
redward by $\sim$ 710 m s$^{-1}$
with respect to the $^7$Li-only profile.
Under ideal conditions, where the observed line shifts are
independent of their strength and carrier, it is straightforward
to determine the  wavelength scale from Fe\,{\sc i} and other lines
and to estimate the $^6$Li contribution from the 6707 \AA\ line's
wavelength.
Convective shifts
in the atmosphere introduce a trend
in the relation between a line's wavelength and
strength (Figure 2). In the case of HD\,82943, the Fe\,{\sc i} lines
of strength $W_{\lambda}$ $\approx$ 50 m\AA\ are shifted by 120 m s$^{-1}$ relative
to lines of strength $W_{\lambda} \sim$ 10 m\AA.
The differential shift is evident from syntheses of the 6707 \AA\
region. When all lines are given an identical radial velocity,
it proved impossible to
fit simultaneously the Li\,{\sc i} line and the adjacent weaker Fe\,{\sc i} at 6707.433
\AA. While the latter line was matched in strength and width, the wavelength
separation between it and the Li\,{\sc i} line was not.
When the stellar radial velocity is measured from the Fe\,{\sc i} lines at 6703~\AA\ and 6705~\AA,
a pair of lines similar in strength to the Li line,
the observed wavelength of the deepest part of the Li\,{\sc i} line is at the predicted wavelength
for $^{6}$Li/$^{7}$Li = 0.0 to within about 1 m\AA. This predicted wavelength is moderately sensitive to small
amounts of $^{6}$Li, for example, the shift from $^{6}$Li/$^{7}$Li = 0.0 to 0.1 is 0.005 m\AA\ ($\sim$ 220~m s$^{-1}$),
but use of the wavelength is compromised by the presence of convective shifts.

As our primary monitor of the lithium isotopic ratio, we use the profile of the 6707~\AA\ line. This
is less affected by the granulation.
The Li abundance and the
$^6$Li/$^7$Li ratios for the program stars are derived with
the following procedure by the UT group. Using the model parameters given in Table~2 and
the  line list given in Table~3, Li-profiles are computed
for a given $^6$Li/$^7$Li ratio by varying the Li abundance and
the wavelength of the Li\,{\sc i} line
 such that the predicted profile
fits best the observed profile. We also took the liberty to 
adjust
the observed profile vertically within
 the rms error ($\sim$ S/N$^{-1}$)
of the continuum.
Abundances of elements (C, N, Ti, V, and Fe)
contributing lines
 that are blended with the Li-profile
are taken from the sources which provided the atmospheric
parameters (see Table~1 for references).
Predicted profiles are first broadened with
the single Gaussian parameter $\Gamma_{g}$
and compared with the observed profile. Then, the  procedure is repeated
but the predicted profiles are broadened using the  combination of
$v\sin i$, $V_m$, and
the instrumental broadening.

\begin{figure*}
\centering
\epsfxsize=18truecm
\epsffile{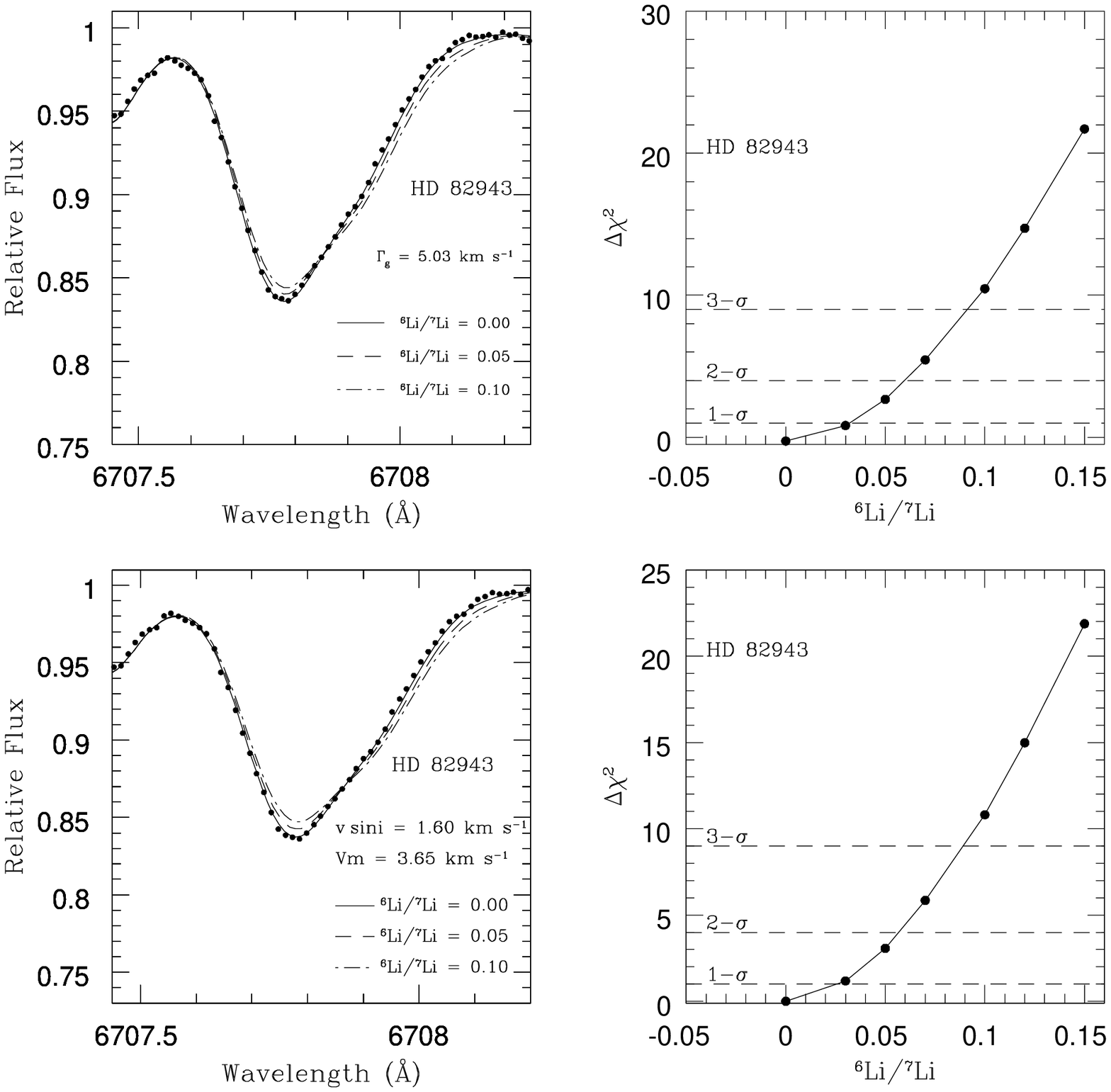}
\caption{ Computed synthetic profiles for different $^{6}$Li/$^{7}$Li
ratios are compared with the observed Li-profile. Computed profiles
are broadened with the broadening parameter $\Gamma_{g}$ (top panel) and
with the separated radial-tangential macroturbulence,
rotational velocity and the instrumental profile (bottom panel).
The $\chi^{2}$ variance $\Delta\chi^{2}$ versus the Li isotopic ratio is shown
for both the cases.
The errors 1-$\sigma$, 2-$\sigma$, and 3-$\sigma$ are noted. }
\end{figure*}

Sample results are shown in Figure 6 for HD 82943 for the
case where rotation and macroturbulence are treated as separate
effects. Synthetic profiles for the case of Gaussian broadening parameter $\Gamma_{g}$ alone
are not discernibly different.
The agreement between the predicted and the observed profiles
is quantified using $\chi^{2}_r$. The minimum values, $\chi^{2}_{min}$,
obtained from $\chi^{2}_{\rm r}$ analysis are given in Table~2. 
In Figure~6, the minimum $\chi^{2}$ variation values, $\Delta\chi^{2}$, 
for HD\,82943
obtained for different $^6$Li/$^7$Li ratios  are plotted.
The $\Delta\chi^{2}$  is computed by subtracting $\chi^{2}_{\rm r}$
from the $\chi^{2}_{\rm f}$; where $\chi^{2}_{\rm f}$ is computed
by dropping wavelength as a free parameter. The wavelength is fixed
for the minimum $\chi^{2}_{min}$ value from the $\chi^{2}_{\rm r}$ analysis.
Thus, by definition both the $\chi^{2}_{\rm r}$ and $\chi^{2}_{\rm f}$ should have the same minimum value
of $\chi^{2}_{min}$.
For HD\,82943, we obtained the  abundance
$\log\epsilon$(Li) = 2.435$\pm$0.005 and the ratio $^6$Li/$^7$Li = 0.00$\pm$0.03.
The quoted 1-$\sigma$ errors in the isotopic ratios (Table~3) are
obtained from the $\chi^{2}$ variation,
the formal
1-$\sigma$, 2-$\sigma$, and 3-$\sigma$ errors
(Bevington \& Robinson 1992) are indicated in Figure~ 6.
The quoted errors in the $\log\epsilon$(Li) are 1-$\sigma$
errors obtained from the $\chi^{2}$ variation of the Li-profile fitting, and
do not include errors arising from the choice of model atmosphere parameters, especially the effective
temperature.

In Table~2, we  give the  results for the lithium abundance and isotopic
ratio  obtained in the UT analysis using
the assumed single line broadening parameter $\Gamma_{g}$.
Also, values of $\chi^{2}_{min}$ for the final best
fit between the observed and the predicted Li-profile are given in Table~2.
The  abundance and the isotopic ratio show very weak
dependence on how the velocity broadening is assigned.
Additional examples of observed and synthetic spectra and accompanying
plots of $\chi^2$ are shown in Figures 7  $\&$ 8.
Although the lithium abundance is sensitive to the adopted
model atmosphere -- temperature is the controlling influence --
the isotopic ratio is quite insensitive to it.
We repeated the
analysis using models drawn from the MARCS grid (Gustafsson et al.
1975)
and found no noticeable
difference in  either  the broadening parameters or the Li-isotopic ratio.
Table 2 shows that, across the sample of stars, the lithium is
pure $^7$Li with no convincing detection of $^6$Li.
This result holds for
stars with and without extra-solar planets.

Our analysis of the 6707 \AA\ line profile using standard model atmospheres
obviously neglects the influence of the stellar granulation, which is
responsible for the line shifts and asymmetries. These
can have only a slight effect on the results. Stellar equivalents of
sunspots and faculae might cause the lithium line to be strengthened
over a few localized regions of a stellar disk. Consider, for example,
a strengthening of the line in an area close to the receding limb of the
star. The result would be a red asymmetry to the lithium line, and
quite possibly a weaker asymmetry would be imposed on the Fe\,{\sc i}
comparison lines used to set the broadening parameters.
In these circumstances, one might overestimate the contribution
of $^{6}$Li.
Similary, if areas on the approaching limb carry undue weight in forming
the profile, the $^{6}$Li contribution could be underestimated. 
If a major portion of the Li\,{\sc i}
line is formed in a few special areas, our theoretical profiles would fail
to fit the observed profiles. 

Our demonstration that excellent fits to the 6707 \AA\ profiles are possible
is a good indication that stellar spots and faculae cannot play an
important role. 
Israelian et al. recognized that a false $^6$Li signal could arise
from the neglect of the effects of
localized regions on the formation of the 6707 \AA\ line. They
advanced cogent reasons - photometric and radial velocity stability -
for rejecting this possibility in the case of HD 82943. The fact that our
observed profiles are very similar to their profile (see below)
is additional evidence that the lithium line is not formed preferentially
in a few areas.     

\begin{figure*}
\centering
\epsfxsize=18truecm
\epsffile{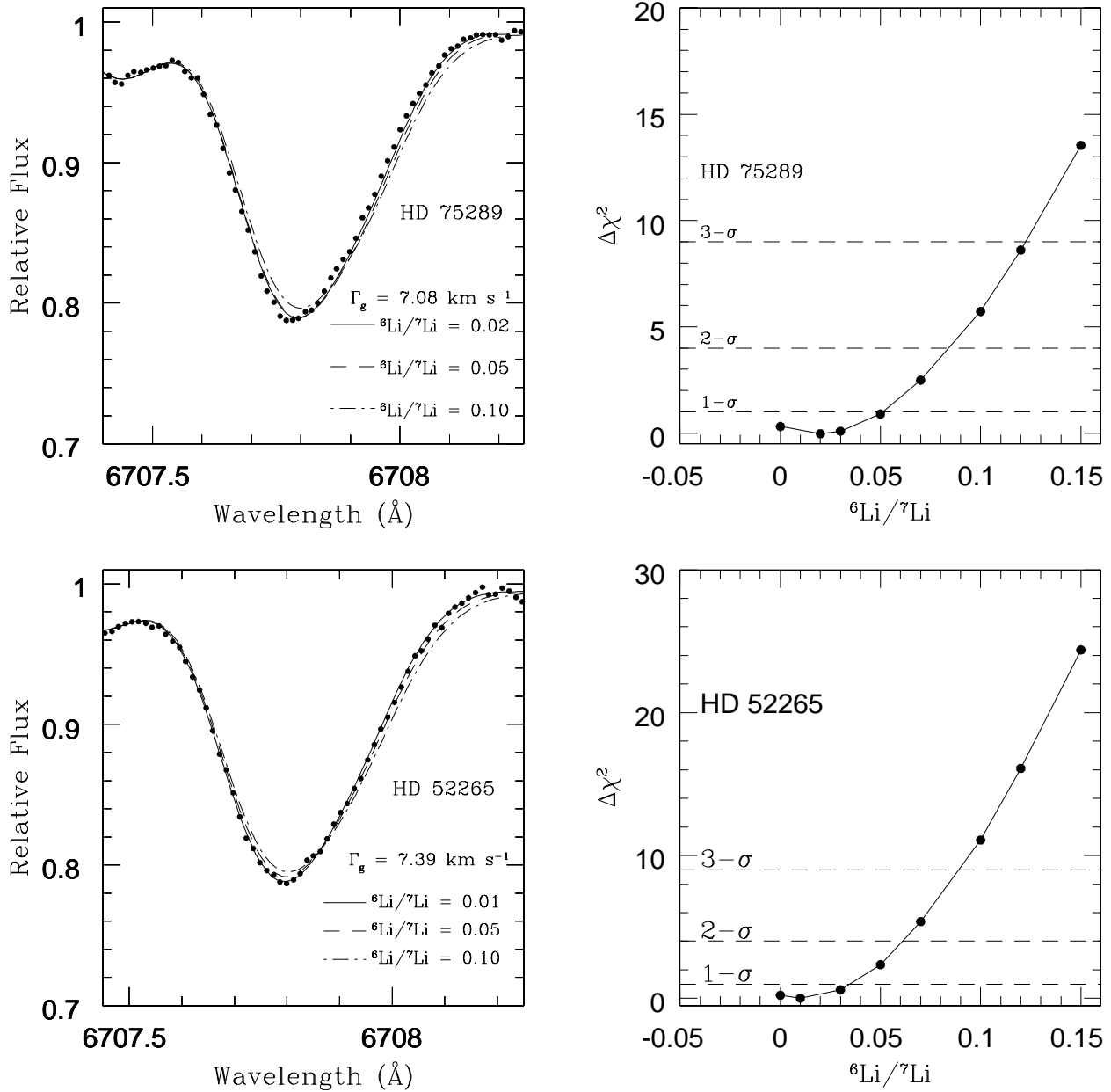}
\caption{ Determination of $^{6}$Li/$^{7}$Li ratio by fitting
the Li-profile for two stars with extra-solar planets: HD\,75289 (top panel) and HD\,52265 (bottom panel).
The $\chi^{2}$-test results are shown for both the stars.}

\end{figure*}

\begin{figure*}
\centering
\epsfxsize=18truecm
\epsffile{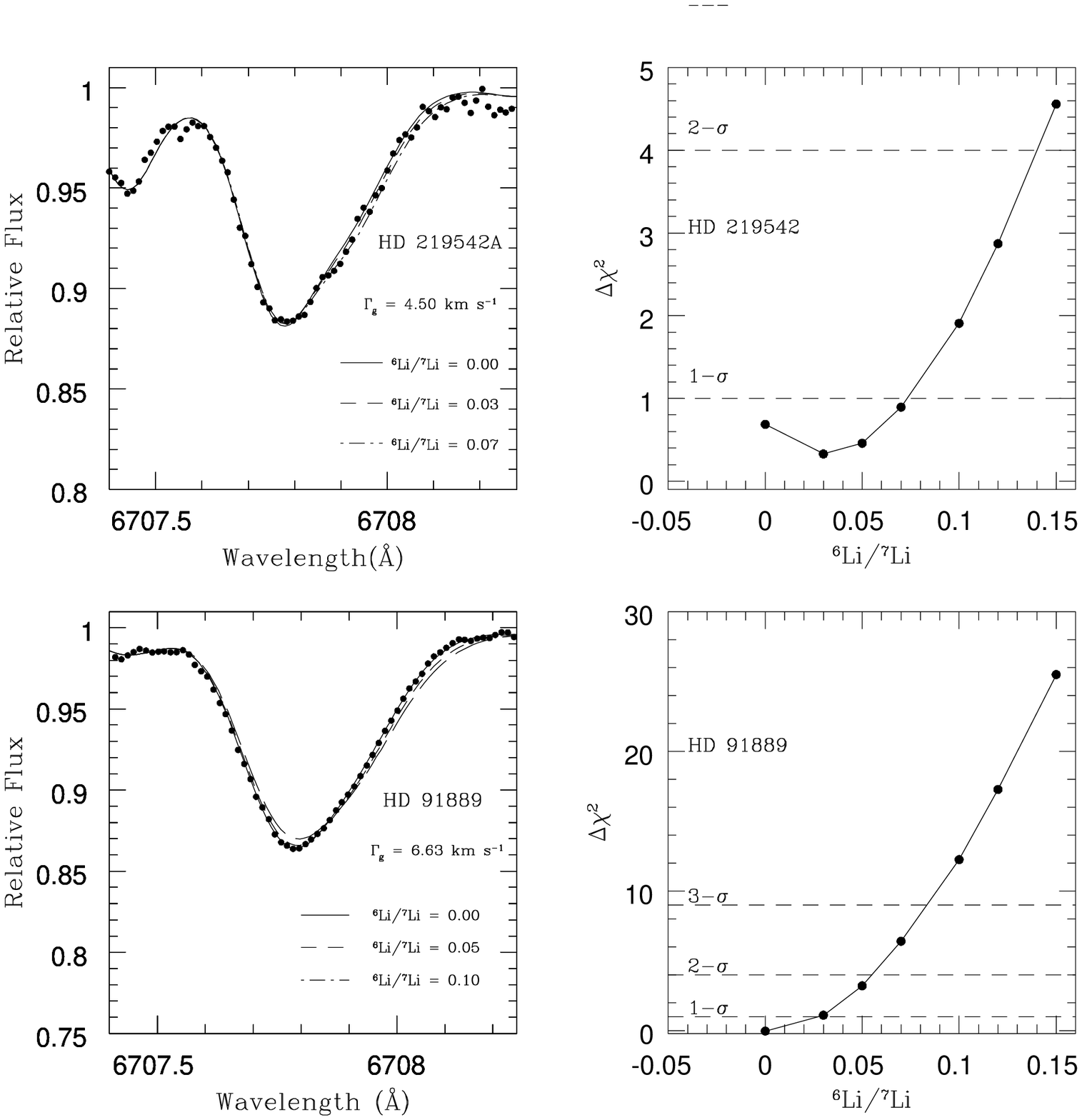}
\caption{ Same as Figure 7. Li-profile fitting and the $\chi^{2}$-test results for
HD\, 219542 A (top panel) and HD\,91889 (bottom panel) which are not known to have planets.
The Li-profile of HD\,219542 best fits the computed Li-profile of $^{6}$Li/$^{7}$Li = 0.03 $\pm$0.04.}
\end{figure*}

\subsection{The Case of HD 82943}

Israelian et al. (2001) derived the high value
of $^{6}$Li/$^{7}$Li = 0.126  $\pm$ 0.014 for HD\,82943.
Our analysis (UT and UW) shows no indication of
$^{6}$Li or $^{6}$Li/$^{7}$Li = 0.0$\pm$0.03.
One can think of two possibilities
for the discrepant results: a) a difference in the observed spectra,
and b) differences in the analyses.

Israelian et al.'s spectrum is of similar quality to ours with respect
to resolving power (R = 110,000 vs our 125,000) and S/N ratio
in the continuum near the Li\,{\sc i} line ( S/N$\sim$ 500 in both the studies).
Comparison of their electronically published Li\,{\sc i} profile for HD\,82943
with our profiles shows near perfect agreement; the differences are much
smaller than those between synthetic profiles for $^{6}$Li/$^{7}$Li = 0.0 and 0.13.

Given that there is  no difference between Israelian et al.'s and our spectra of a size
sufficient to account for the different $^6$Li/$^7$Li ratios, the
sources of the discrepancy must lie in the methods of analysis.
In seeking an explanation, we consider the adopted model atmospheres, the fitted line
broadening parameters, and the chosen line lists.
We adopt the model atmosphere parameters -- $T_{\rm eff}$,
$\log g$, [Fe/H], and $\xi_{\rm t}$ --  used by Israelian et al.
They do not explicitly indicate the source of the model atmosphere
but it is most improbable that given the similarities between modern
grids of model atmospheres (see our test of a MARCS model) that
an alternative choice can lead to anything but a very small change in the
recovered $^6$Li/$^7$Li ratio. 

In obtaining the broadening parameters,
Israelian et al. used two of our three Fe\,{\sc i} lines to obtain
$v\sin i$ = 1.65 $\pm$ 0.05 km s$^{-1}$ and $V_m$ = 3.90 $\pm$ 0.2 km s$^{-1}$.
These compare well with our values $v\sin i$ = 1.60 $\pm$ 0.20 km s$^{-1}$,
and $V_m = 3.65 \pm$ 0.10 km s$^{-1}$, where we assumed the
same form for $V_m$ as Israelian et al. Analysis
of our spectrum using their broadening parameters also gives
$^6$Li/$^7$Li = 0.01, not 0.126. Our result is insensitive to the form adopted
(Gaussian macroturbulence vs rotation and macroturbulence) for the line broadening (Figure 9).

\begin{figure*}
\centering
\epsfxsize=18truecm
\epsffile{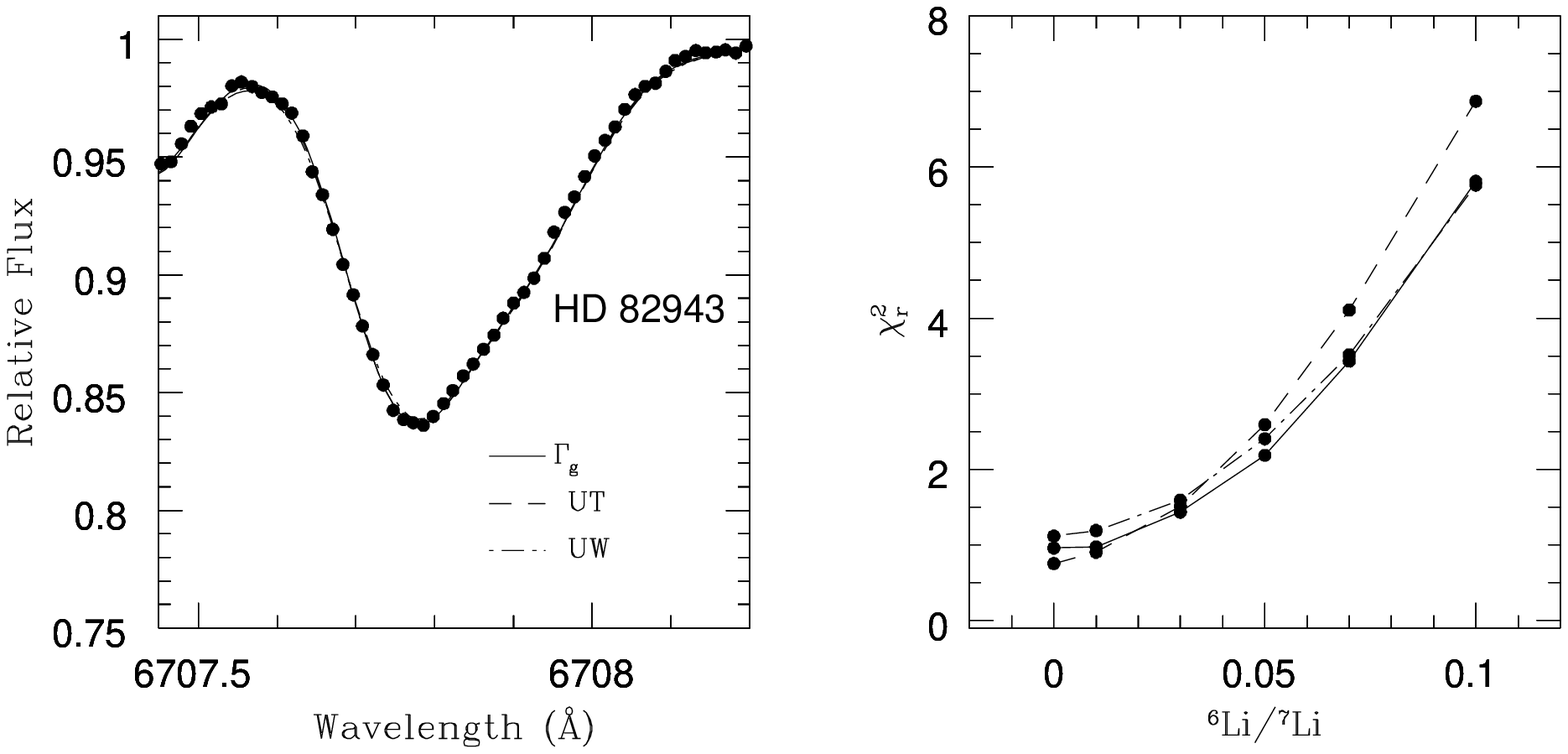}
\caption{ Li-profile fitting for HD\,82943 using three different forms
of line broadening parameters (left panel): $\Gamma_{g}$~=~5.03 km s$^{-1}$), 
v\,sini = 1.60 km s$^{-1}$, V$_{\rm m}$ = 3.65 km s$^{-1}$
from the UT analysis and
v\,sini = 2.09 km s$^{-1}$, V$_{\rm m}$ = 3.73 km s$^{-1}$ from the UW analysis.
In the right panel we showed
$\chi^{2}_{\rm r}$-test analysis for each case. Note a very slight change in the quality of
the fit but in all the three cases $\chi^{2}_{\rm min}$ is found for $^{6}$Li/$^{7}$Li $\leq$ 0.01.}
\end{figure*}

The inferred contribution from $^6$Li to the observed profile
depends on the completeness and accuracy of the line list.
It would appear from their paper that Israelian et al. did not include the
unidentified line at 6708.025 \AA\ in their line list, and overlooked
the weaker line at 6708.125 \AA.
The unidentified lines which we assign to Ti\,{\sc i}
play a role because they lie on either side of the weaker $^6$Li\,{\sc i}
line. Their combined strength in the solar spectrum is about 1.5 m\AA. In a metal-rich
star like HD 82943, their strength is greater than in the Sun. 
Given that the
equivalent width of the 6707 \AA\ feature is 50 m\AA\ for HD\,82943, the
lines' estimated contribution
of 2.2~m\AA\ to the weaker $^6$Li line corresponds to $^6$Li/$^7$Li $\sim$ 0.06.
\begin{figure*}
\centering
\epsfxsize=16truecm
\epsffile{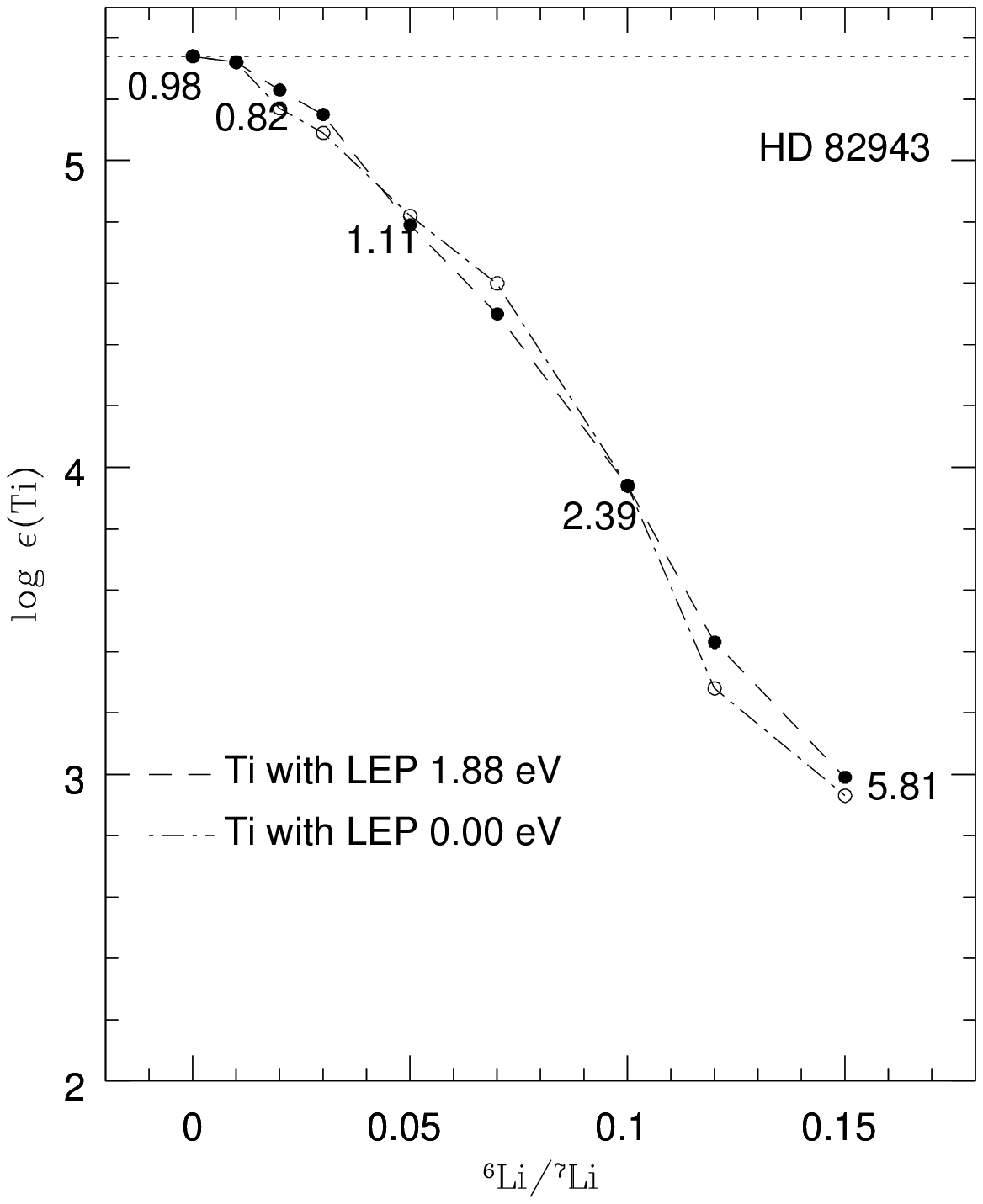}
\caption{ Investigation of the effect of adopted Ti\,{\sc i} abundance on the
$^{6}$Li/$^{7}$Li ratio for HD\,82943. The dotted horizontal
line is the observed Ti\,{\sc i} abundance for HD\,82943. The numbers
are $\chi^{2}$-values for the best fits for 1.88 eV case. }
\end{figure*}
The linelist compiled independently at the University of Washington differs very little
from that in Table~3. Use of this linelist instead of that in Table~3 does not give
significantly different synthetic spectra computed for the same input data. 
Israelian et al. used HD\,91889 as a comparison star, but this is not necessarily a perfect
foil for eliminating the effects of weak blends. This star is metal-poor with [Fe/H]~=~$-$0.23 or
0.55~dex down from HD\,82943. This relative metal deficiency plus slight differences in effective
temperature and surface gravity mean that unidentified lines like the supposed Ti\,{\sc i} lines
will be appreciably weaker in HD\,91889 than in HD\,82943. In contrast, our sample of comparison stars
includes examples of similar metallicity to HD\,82943.

To investigate further the influence of the unidentified lines, we analysed the HD\,82943 profile
with different Ti abundances assigned to the pair of unidentified lines. As the abundance
is decreased, the drop in absorption from the Ti\,{\sc i} lines is compensated for an 
increasing $^{6}$Li abundance, but the quality of the fit to the observed line
decreases steadily. This is well shown by Figure 10 where we plot the Ti abundance
versus the derived $^{6}$Li/$^{7}$Li that best fits the line. The exercise was done
for Ti\,{\sc i} lines of 0.0 eV and 1.88 eV. In the latter case, the $\chi^{2}$ for the best
fit is marked and clearly increases as the contribution of the unidentified lines is reduced.

Our stellar sample include several other metal-rich stars, including one
not known to host planets. A plausible assumption is that, if $^{6}$Li is
present in these stellar atmospheres, the $^{6}$Li/$^{7}$Li ratio will
vary from star-to-star, and is likely absent in the star without
planets (HD\,75332). Incorrect representation of the list of blending
lines could bias the $^{6}$Li/$^{7}$Li determinations but is unlikely
across this sample of metal-rich stars to alter the spread in $^{6}$Li/$^{7}$Li ratios.
The fact that we find $^{6}$Li/$^{7}$Li $\simeq$ 0.00 in all cases suggests that $^{6}$Li
is not present in these metal-rich stars, including HD\,82943. 
Blends have a smaller effect on the more metal-poor stars.
The best fits to the observed profiles are found for $^{6}$Li/$^{7}$Li = 0.00  (i.e., non-negative) is a
good indication that the blending lines have not been overestimated.

\section{Discussion}
Our sample of planet-hosting stars are placed in an HR-diagram (Figure 11)
as a guide to their evolutionary state.
We use  the {\it Hipparcos}
parallaxes, the observed
$V$ magnitudes (interstellar reddening is neglected), and the reported
$T_{\rm eff}$.  On this diagram, we place isochrones from Girardi et al. (2000)
for [Fe/H] = 0.0 and 0.3 and ages from 1 to 9 Gyr.  All but three of the stars
are young; their ages are 4 Gyr or less (Table 1).

\begin{figure*}
\centering
\epsfxsize=12truecm
\epsffile{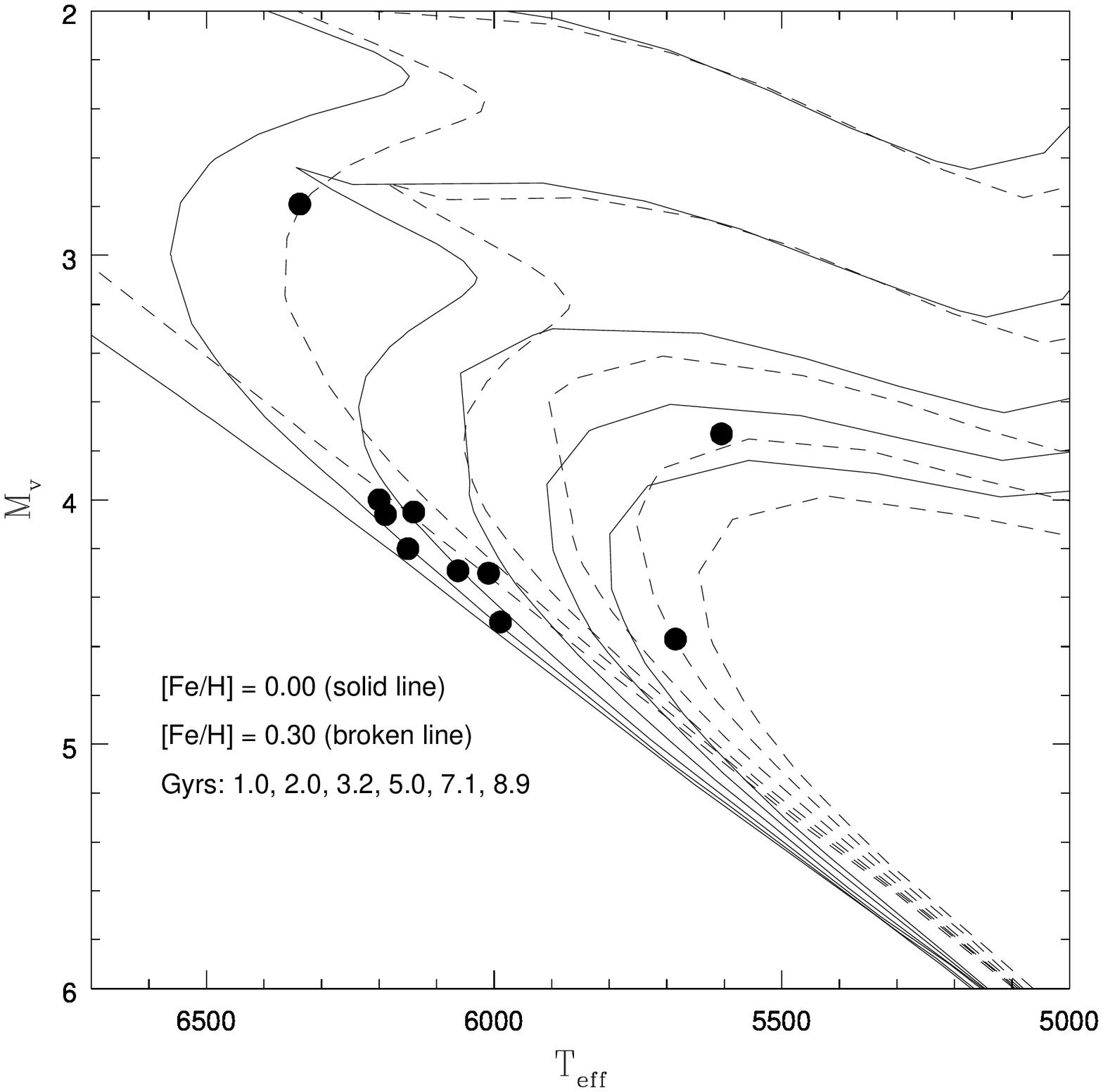}
\caption{ Age estimation using the isochrones
computed by Girardi et al (2000) for program stars with planets. }
\end{figure*}

The  young stars have a lithium ($^7$Li) abundance that appears normal for stars of
their age and metallicity.  An empirical demonstration of normality is provided
in
Figure 12 where the lithium abundance is plotted versus $T_{\rm eff}$
and compared with observations from three open clusters of differing
ages -- Pleiades at 0.1 Gyr, Hyades at 0.7 Gyr, and M67 at 4.5 Gyr --
summarised by Jones, Fischer, \& Soderblom (1999). Ages inferred from Figure 11
are consistent with those listed in Table 1.  Lithium abundances for our main
sequence stars and the clusters are approximately consistent with theoretical
predictions (see, for example, Proffitt \& Michaud 1989).
We note that our sample stars were selected to have moderate to strong Li lines.
So, the plot is necessarily biased towards higher Li abundances.

\begin{figure*}
\centering
\epsfxsize=12truecm
\epsffile{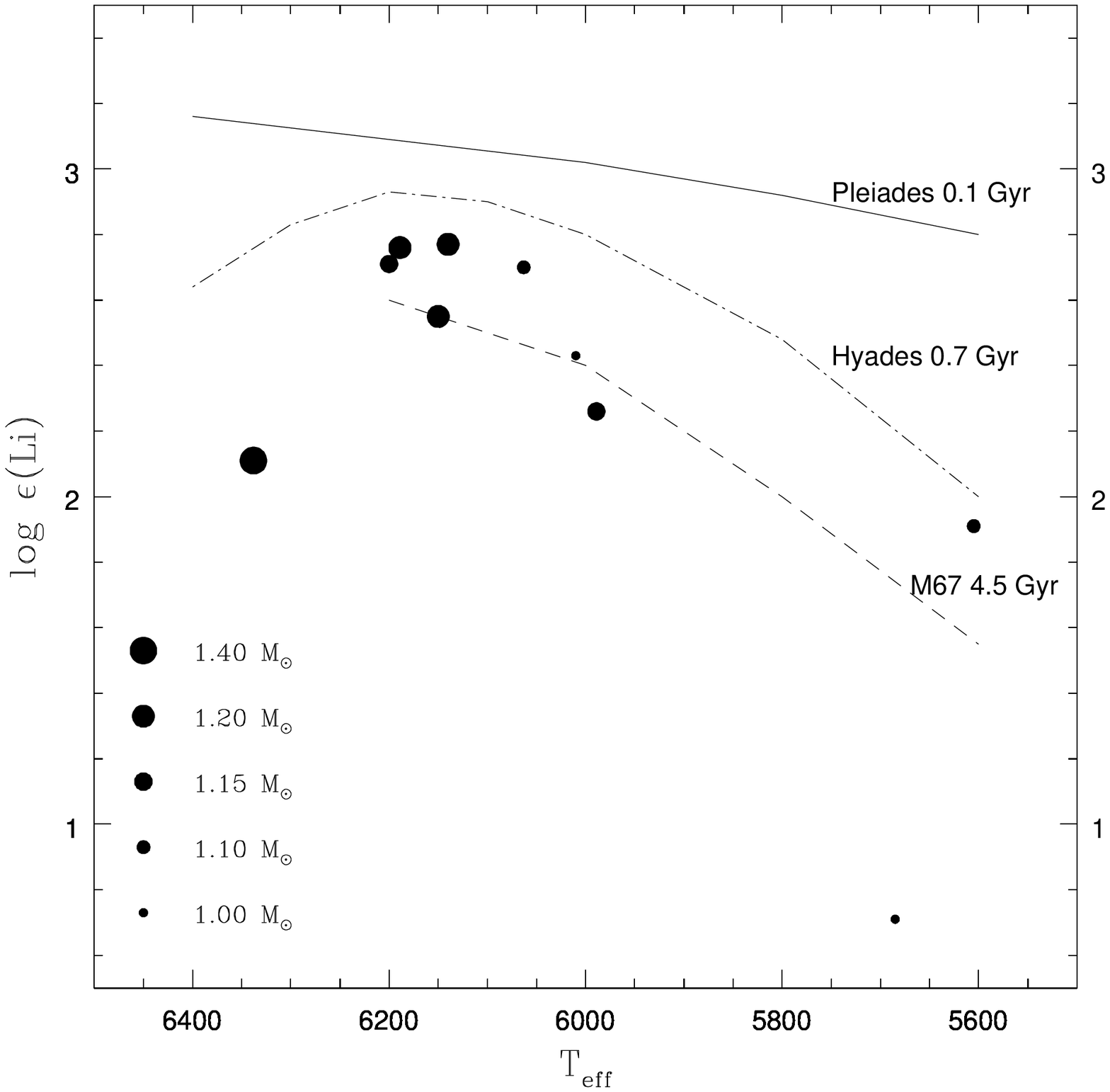}
\caption{ The derived Li abundances for planet hosting stars
are compared with the Li abundances of Pleiades, Hyades and M67 clusters. Ages
of the clusters are noted. The stellar masses of planet hosting stars are indicated by the
size of the filled circles.}
\end{figure*}

Normal behaviour for $^6$Li must be defined on theoretical grounds because
there are no observations of this isotope in young stars. 
Standard models (Proffitt
\& Michaud 1989) predict effectively complete destruction of $^6$Li in the
pre-main sequence phase. Destruction results when the
convective envelope exposes lithium to warm protons at its base. Little to no
destruction of $^7$Li occurs;  the ($p,\alpha$) reactions occur about 80
times more slowly for
$^7$Li than for
$^6$Li.  Destruction of  $^7$Li occurs on the main sequence, as revealed by the
lowering of the lithium abundance in  the older clusters. Loss of lithium is
more severe the lower the mass of the star. Since the lithium depletion is
driven by exposure to protons, even  minor reductions of the $^7$Li
abundance below the value in zero age main sequence stars must be accompanied by
very much larger reductions for
$^6$Li.

Into this picture for lithium abundances of normal stars must be woven the
suggestion that main sequence stars may accrete one or more planets (or
circumstellar material) which provide fresh lithium to the star's convective
envelope with the largest effect being a replenishment of the $^{6}$Li abundance.
After accretion, the $^6$Li/$^7$Li ratio in the star's atmosphere
depends primarily on (i) the mass and composition of ingested
material relative to the mass and composition of the stellar convective
envelope, and (ii) the time elapsed since ingestion.
Montalb\'{a}n \& Rebolo (2002)
have provided a useful exploration of these dependencies. Their calculations for
 main sequence stars consider the accretion of 1 to 10 $M_J$
by main sequence stars of masses from about 0.7 $M_\odot$ to 1.2 $M_\odot$ and of
two compositions ([Fe/H] = 0.0 and 0.3). A ratio $^6$Li/$^7$Li $\simeq$ 0.10
immediately following accretion is achieved, for example, when a giant planet of
mass 10
$M_J$ is accreted by a 1.2 $M_\odot$ main sequence. The post-accretion isotopic
ratio decreases with decreasing mass of the star because the mass of a main
sequence star's convective envelope increases to lower masses.
 Accretion increases also
the $^7$Li abundance, that is the `total' lithium abundance. If lithium has not
been depleted prior to accretion, the abundance increase is small (about 0.2
dex for accretion levels providing $^6$Li/$^7$Li $\sim$ 0.10), and, therefore,
the total lithium abundance is not a sensitive indicator of the occurrence of
accretion.\footnote{Similarly, the beryllium abundance is a poor sensor of
accretion. In fact, beryllium is largely undepleted in main sequence stars,
even in stars showing lithium to be severely depleted. Santos et al. (2002)
measured Be abundances to find `no clear difference' between the abundances in
stars with and without planets.}

Survival of
$^6$Li also  determines its detectability.
Within the constraints set by plausible
assumptions about the accretion process and its benign effects on the
structure of the star, the calculations show that accreted $^6$Li (and $^7$Li)
is preserved for more than 3 Gyr in stars of mass greater than about 1.1
$M_\odot$. This limit corresponds to the effective temperature range $T_{\rm
eff} \geq 5900$ K. Lithium-7 is predicted to be completely preserved, a
prediction at odds with the lithium abundances  for field and cluster stars
(Figure 12). This discrepancy could be removed by using alternative models, see,
for example, the predictions given by Proffitt \& Michaud (1989). One can infer
from observations summarized in Figure 12 that accreted $^6$Li will be depleted
more quickly than predicted by Montalb\'{a}n \& Rebolo's standard calculations:
$^6$Li seems likely to survive for only about 1 Gyr in main sequence stars with
$T_{\rm eff} \geq$ 6100 K.
(For $T_{\rm eff} \sim 6600$ K, lithium is
depleted to form the Boesgaard-dip [Boesgaard \& Tripicco 1986]), but
both isotopes may be quite similarly depleted.)

In the light of Figure 12 and Montalb\'{a}n \& Rebolo's calculations, our
failure to detect $^6$Li in the young planet-hosting main sequence stars does
not admit of a simple decisive conclusion about accretion of planets.
Our measurements allow the following
possibilities: (i) accretion has not occurred; 
(ii) accretion occurred but the
accreted mass was insufficient to raise the star's $^6$Li abundance above our
detection limit; and (iii) accretion occurred such that  the post-accretion $^6$Li
abundance exceeded the detection limit but subsequent destruction of lithium
depressed the $^6$Li abundance again below the detection limit. Montalb\'{a}n \&
Rebolo's predictions of the survivability of $^6$Li in stars with $M \geq 1.1
M_\odot$ and our failure to see $^6$Li may suggest that accretion can rarely
add more than a Jovian mass of `primordial' material to the star.  (Accretion of
terrestrial rather than Jovian planets can effect an increase in the lithium abundances, but
with a much smaller total mass for the accreted planets.)

Effects of accretion of planets on surface compositions are not limited to the
lithium abundance.
Accretion of circumstellar material has been invoked to account for the
metallicity of planet-harbouring stars. There is now a  general concensus that
such stars are more metal-rich  than similar stars without
planets (Gonzalez et al. 2001; Santos et al. 2001; Reid 2002), but the origin
of this difference is still a matter of debate. Gonzalez (1997) suggested two
possible explanations for the metallicity difference: accretion of H-poor matter
(e.g., terrestrial planets, asteroids), and/or a bias in the discovery of
planets in radial velocity surveys arising from a metallicity-dependent
migration of giant planets toward the star. Gonzalez (1998) added a third
possibility: a high metallicity of the natal cloud as a prerequisite for
formation of giant planets. Santos et al. (2001) favoured this latter
possibility. Combinations of these possibilities may need to be considered; for
example,  if high metallicity clouds are favoured sites for stars with planets,
such stars may accrete  planets, and if these planets are terrestrial
in nature, the metallicity of the star will be further increased.

Observational evidence in support of accretion as the origin of the high
metallicity of planet-hosting stars is limited. Detection of $^6$Li in HD
82943 was claimed as supporting evidence. Our failure to confirm the presence of
$^6$Li weakens this claim, but, as we noted, accretion remains a
possibility as long as $^6$Li can be efficiently destroyed after the
cessation of accretion.
Accretion of giant planets will probably not
change the surface composition of a star very much; giant planets are expected to have a
composition similar to that of their host star, except for lithium and,
perhaps, beryllium and boron.
 Since the composition of a terrestrial planet is far
from a replica of the circumstellar material, the stellar composition will be
changed by accretion of terrestrial planets.
  One expects volatile elements (e.g., C, N, and O) to be
reduced, and the abundance of  non-volatiles (e.g., Fe-peak elements)
to be increased by accretion. A signature of accretion of terrestrial planets
could be an abundance change correlated with an element's condensation
temperature ($T_c$), the temperature at which an element condenses into solids
as circumstellar gas  cools in thermodynamic equilibrium.
Gonzalez (1997) proposed this signature and searched for it  in abundance
analyses of two planet-hosting stars. Smith, Cunha,
\& Lazzaro (2001) noted that single (planet-less) metal-rich stars have a
composition such that abundance differences, [X/H] (relative to solar values) for element (X) are
quite well correlated with  $T_c$. These differences are, therefore, plausibly attributed to
Galactic chemical evolution, that is the net effect of nucleosynthesis by stars on the chemical composition of the
interstellar medium.
Stars hosting planets show the same abundance -- $T_c$ correlation at a given [Fe/H] indicating
that accretion of terrestrial planets can have more than a minor role in
setting the stars' surface composition (see Takeda et al. 2001). Smith et
al. did draw attention to a sample of five planet-hosting metal-rich stars  for which the [X/H] versus
$T_c$ correlation appeared unusually strong and suggested that accretion  of
terrestrial planets might have occurred in these cases. Four of the five stars
are in Table 1. 

Another example of a star which may have accreted dust, rocks, or terrestrial-like planets
is HD\,219542~A, a main sequence star with a distant main sequence companion. Gratton et al.'s (2001)
differential abundance analysis found HD\,219542~A overabundant relative to HD\,219542~B
in non-volatile elements but not in the volatile elements. The former were overabundant by about
0.07~dex. Gratton et al. remark that accretion of a few earth masses of rocky material by HD\,219542~A
would suffice to account for the differential abundances, and might also explain the different
lithium abundances in primary and secondary; lithium's condensation temperature is similar
to that of silicon and iron. HD\,219542~A is one of our programme stars and the Li abundance (Table~2)
agrees well with that given by Gratton et al. (HD\,219542~B is Li-poor, $\log\epsilon$ (Li) $<$ 1.0,
according to Gratton et al.)
The lithium abundance of HD\,219542~A is 'normal' for this young star.

If accretion occured and added lithium, the $^{6}$Li abundance corresponding to a 0.07~dex
increase in the iron abundance would be approximately $\log\epsilon$($^{6}$Li) = $-$0.1,
if the accreted material contained a `cosmic' abundance of lithium relative to iron. The
corresponding $^{7}$Li abundance is about $\log\epsilon$($^{7}$Li) = 0.9 in the
absence of lithium at the stellar surface prior to accretion.
The observed abundance is considerably higher implying, as expected from Figure~12, that
substantial amounts of $^{7}$Li were present when accretion occured. Combining the
observed $^{7}$Li abundance with the inferred $^{6}$Li abundance from the mass of accreted rocky
material gives $^{6}$Li/$^{7}$Li = 0.04. We measure $^{6}$Li/$^{7}$Li = 0.03~$\pm$ 0.04 which
is consistent with the inference but also with an absence of $^{6}$Li. 
Inspection of the observed profile gives a tantalising impression
of a break in the red wing due to $^{6}$Li (Figure~7). Higher S/N spectra will be sought.

\section{Concluding Remarks}

Our search for $^6$Li in the atmosphere of planet-hosting star proved
unsuccessful, despite the intriguing earlier report of substantial amounts of
$^6$Li in HD 82943, a star known to host two giant planets. A limit of
approximately $^6$Li/$^7$Li $\leq$ 0.03 was set for eight planet-hosting stars
including HD 82943. Detection of $^6$Li in a late F/early G-dwarf was recognized by Israelian et al.
(2001) as a possible signature of accretion of planets by a star.
Non-detection of $^6$Li is less than watertight evidence against such
accretion. 
Searches
for planets should in the near future identify much larger numbers of  planet-hosting
stars. Accretion of circumstellar material may also occur onto  stars without a
planetary system. A continued search for $^6$Li may yield a positive detection of $^{6}$Li.

\section*{Acknowledgments}
We thank Carlos Allende Prieto, Gajendra Pandey, and 
Nils Ryde, for many spirited discussions, and UW group thanks Verne Smith for offering software.
We thank Anita Cochran, Bill Cochran, Garek Israelian and Rafael Rebolo for their comments. 
The research of the UT group has been supported in part by National Science Foundation (grant AST 96-18414)
and the Robert A. Welch Foundation of Houston, Texas. The UW group's research has been supported
by a NASA Astrobiology Institute grant.
This research has made use of the SIMBAD
data base, operated at CDS, Strasbourg, France, and the NASA ADS service, USA.

\end{document}